\documentclass[aip, preprint, pop]{revtex4-1} 
\usepackage{times}		
\usepackage{booktabs}
\usepackage{placeins}

\usepackage{amssymb,amstext,amsmath, epsf}
\usepackage{graphicx,graphics,epsfig,wrapfig}
\usepackage{listings}
\usepackage{float}
\usepackage{esint}

\usepackage{color}
\usepackage[english]{babel}
\usepackage[T1]{fontenc}
\usepackage[utf8]{inputenc}
\usepackage[font=small,labelfont=bf]{caption}
\usepackage{mwe}    
\usepackage{subcaption}
\usepackage[normalem]{ulem}

\newcommand{\bearr}{\begin{eqnarray}}
\newcommand{\eearr}{\end{eqnarray}}

\newcommand{\V}[1]{\bm{#1}}

\newcommand{\unitvec}[1]{\widehat{\V{#1}}}
\newcommand{\be}{\begin{equation}}
\newcommand{\ee}{\end{equation}}

\newcommand{\micronbase}{$\mu$m}
\newcommand{\micron}{\micronbase}
\newcommand{\microns}{\micronbase\ }

\newcommand{\Zeff}{Z_{\mathrm{eff}}}

\usepackage[normalem]{ulem}

\begin{document}

\title{Hybrid-VPIC: an Open-Source Kinetic/Fluid Hybrid Particle-in-Cell Code}

\author{Ari Le}

\affiliation{Los Alamos National Laboratory, Los Alamos, NM 87545, USA}
\author{Adam Stanier}
\affiliation{Los Alamos National Laboratory, Los Alamos, NM 87545, USA}
\author{Lin Yin}
\affiliation{Los Alamos National Laboratory, Los Alamos, NM 87545, USA}
\author{Blake Wetherton}
\affiliation{Los Alamos National Laboratory, Los Alamos, NM 87545, USA}
\author{Brett Keenan}
\affiliation{Los Alamos National Laboratory, Los Alamos, NM 87545, USA}
\author{Brian Albright}
\affiliation{Los Alamos National Laboratory, Los Alamos, NM 87545, USA}

\date{\today}

\begin{abstract}
Hybrid-VPIC is an extension of the open-source high-performance particle-in-cell (PIC) code VPIC incorporating hybrid kinetic ion/fluid electron solvers. This paper describes the models that are available in the code and gives an overview of applications of the code to space and laboratory plasma physics problems. Particular choices in how the hybrid solvers were implemented are documented for reference by users. A few solutions for handling numerical complications particular to hybrid codes are also described. Special emphasis is given to the computationally taxing problem of modeling mix in collisional high-energy-density regimes, for which more accurate electron fluid transport coefficients have been implemented for the first time in a hybrid PIC code.
\end{abstract}

\maketitle

\section{Introduction}
\label{s:intro}

In this paper, we present the Hybrid-VPIC code, which is a general-purpose kinetic ion/fluid electron hybrid version of the high-performance particle-in-cell (PIC) code VPIC \cite{bowers:2008}. Hybrid-VPIC was developed to leverage the excellent performance of the VPIC code\cite{bowers:2008b,bowers:2009} as well as its flexibility in allowing users to specify new problems, boundary conditions, and custom physics modules. The Hybrid-VPIC code is part of the open-source VPIC code\cite{hybridvpic}, and a main goal of this paper is to serve as a reference for others who would like to use the code.

In general, hybrid codes\cite{lipatov:2002,winske:2003,winske:2022} allow the modeling of systems where one species (typically the electrons) can be treated as a fluid, while other species (usually ions) require a kinetic treatment. The primary advantage of hybrid models is that they do not need to resolve the smallest kinetic scales in the system if those scales are deemed unimportant to the physics problem at hand. This can reduce computational costs by orders of magnitude. For typical space physics applications, for example, neglecting the electron kinetic scales typically reduces the computing cost and memory requirements by a factor on the order of $\sqrt{m_i/m_e}$ ($m_i/m_e$ is the ion-to-electron mass ratio) for each spatial dimension and the time step may usually be increased by a similar factor. Hybrid codes thus allow relatively large-scale dynamics to be captured while retaining ion kinetics, and they form a bridge between fully kinetic models and fluid models.

A number of hybrid PIC codes exist and have been used to study space, astrophysics, and laboratory plasma problems \cite{nieter:2004,gargate:2007,karimabadi:2011,muller:2011,omelchenko:2012,kunz:2014,fatemi:2017,muller:2011,peterson:2018,haggerty:2019,cohen:2019}. The core hybrid algorithm used in Hybrid-VPIC is largely the same as that used in the code H3D \cite{karimabadi:2011}, which was previously run on a number of petascale parallel computers. The algorithm uses explicit time-stepping, making it relatively simple and efficient. This has allowed Hybrid-VPIC to achieve good performance (around 10 million particle pushes/second/processor on CPU-based computers) and scale up to large problems requiring $>10^5$ processing cores.

The Hybrid-VPIC code is currently being used to explore several different application areas, and some examples are given to demonstrate where the code may be useful. Applications of the code include magnetosphere modeling, other space physics applications, and simulating laboratory fusion devices. Most of the descriptions of applications are kept brief, serving only to illustrate how Hybrid-VPIC can be set up to model a variety of problems and to highlight a couple of numerical complications with hybrid PIC methods. A simplified electrostatic version of the code has been used to model mix in mutli-dimensional flows in high-energy-density (HED) regimes. This application area has required substantial computing resources, and a somewhat more in-depth description of initial science results is given.

The paper is organized as follows: Section~\ref{sec:hybridem} gives a brief review of the equations solved in hybrid models with kinetic ions and massless fluid electrons and describes several particular choices made in implementing the Hybrid-VPIC solver. Also included is an example of open boundary conditions used in electromagnetic Hybrid-VPIC simulations, which may be useful in other hybrid PIC codes. Section~\ref{sec:emapplications} presents a few example problems where Hybrid-VPIC has already been applied. The examples illustrate some numerical complications specific to hybrid PIC models. Section~\ref{sec:hybrides} introduces a simplified electrostatic version of Hybrid-VPIC, which has served primarily to study kinetic mix effects in high-energy density (HED) regimes, an application described in Section~\ref{sec:interface}. A summary discussion follows in Section~\ref{sec:summary}. Section~\ref{sec:ebeam} of the Appendix describes an extension of the code to relativistic electron beam propagation in air, which requires a hybrid model that retains a kinetic treatment of the beam electron population. Details on the particle shapes are given in Sec.~\ref{sec:qsappend}.

\section{Electromagnetic Hybrid Model}
\label{sec:hybridem} 

This section briefly reviews the equations modeled by Hybrid-VPIC along with specific choices made in implementing a hybrid solver in VPIC. The equations and methods are fairly standard for hybrid PIC simulation, and only a few aspects of PIC implementation are specific to Hybrid-VPIC. Some details about particle shapes and boundary conditions are covered, and they may be useful for other hybrid PIC codes.

\subsection{Hybrid model equations}
Hybrid models combine a kinetic treatment of some species with a fluid description of others. Hybrid-VPIC is used primarily as a kinetic ion/fluid electron code similar to a number of other hybrid PIC codes \cite{lipatov:2002,winske:2003}. The model is appropriate for phenomena that have characteristic length scales much larger than the electron kinetic scales (Debye length and electron skin depth) and time scales longer than the electron plasma period. The hybrid model equations are taken in a low-frequency limit that assumes the plasma is quasi-neutral. Each species $s$ of ion of charge $q_s = Z_se$ and mass $m_s$ is described by a Vlasov equation for its phase space distribution $f_s({\bf{x}},{\bf{v}},t)$ of the form:
\begin{equation}
  \frac{\partial f_s}{\partial t} + {\bf{v}}\cdot\nabla f_s + \frac{q_s}{m_s}({\bf{E}} + {\bf{v}}\times{\bf{B}})\cdot\nabla_v f_s  = C_i\{f_s\},   
\end{equation}
where $\bf{E}$ is the electric field, $\bf{B}$ is the magnetic field, and $C_i\{f_s\}$ is an operator that accounts for particle collisions. As in other PIC codes, these equations are solved by tracing sample ion macro-particles along the single-particle trajectories in the electromagnetic fields.

The electron fluid enters the evolution equations through an Ohm's law. We treat the electrons as a massless fluid \cite{winske:2022}, and their momentum balance equation yields the following Ohm's law for the electric field:
\begin{equation}
{\bf{E}} = - \frac{1}{en_e}\nabla p_e  - {\bf{u_i\times B}} +  \frac{1}{en_e}{\bf{J \times B}} + {\bf{R}}_{ei} 
\label{eq:ohm}
\end{equation}
The bulk electron flow is replaced based on ${\bf{J}} = n_ee({\bf{\bar{u}}}_i - {\bf{u}}_e)$, where ${\bf{\bar{u}}}_i = \sum_s q_s \int {\bf{v}} f_s({\bf{v}})  d{\bf{v}}/en_e$ and quasi-neutrality is assumed so that $n_e = \bar{Z}_in_i = \sum_s q_s \int f_s({\bf{v}})  d{\bf{v}}$. Because the hybrid model includes terms proportional to $1/n_e$, a modification is necessary for low-density and vacuum regions. We use the simplest method of applying a density floor $n_f$ in the field solver, so that the density used to advance the fields is $n_e = \max(n_e,n_f)$. Typical values used for the density floor are $n_f/n_0\sim$ 0.01 - 0.05, where $n_0$ is a reference background density. VPIC and Hybrid-VPIC allow different code units to be used, and the reference density $n_0$ is typically the value used to define lengths in terms of the electron skin depth $d_e$ (for "natural" relativistic PIC units in fully kinetic systems with velocities normalized to the speed of light $C$)  or the ion skin depth $d_i$ (convenient for magnetized hybrid simulations using Aflvenic units with velocities normalized to a reference Alfven speed $v_A$).

For electromagnetic problems, we include resistive and hyper-resistive terms of the form:
\begin{equation}
{\bf{R_{ei}}} = \eta{\bf{J}} - \eta_H \nabla^2{\bf{J}},
\end{equation}
where the resistivity $\eta$ may be thought of as accounting in a simple way for electron-ion collisions, and the hyper-resistivity $\eta_H$ may represent an electron viscosity and helps with numerical stability by damping grid-scale oscillations. As described later in Sec.~\ref{sec:hybrides}, the term ${\bf{R_{ei}}}$ may also include contributions from explicit friction or other momentum exchange terms in the particle collision models. While we write Eq.~\ref{eq:ohm} with a simple scalar electron pressure $p_e$, it is possible to incorporate electron pressure tensor effects\cite{le:2016ani}. Hybrid-VPIC is set up to accommodate a full electron pressure tensor, though anisotropic pressure models have not yet been implemented. Note that in the absence of a resistive term ${\bf{R}}_{ei}$ and a pressure gradient, the magnetic field is frozen into the electron fluid in the massless electron hybrid model. The total current density in the Hall term $(1/en_e){\bf{J \times B}}$ is taken from the low-frequency limit of Ampere's law:
\begin{equation}
\mu_0{\bf{J}} = \nabla \times {\bf{B}} 
\end{equation}
The magnetic field is evolved with Faraday's law:
\begin{equation}
\frac{\partial{\bf{B}}}{\partial t} = -\nabla \times {\bf{E}}.
\end{equation}
The Hall term supports Whistler waves, which usually place the strongest limit on the time step in hybrid PIC codes. The Courant–Friedrichs–Lewy (CFL) condition for Whistler waves on the time step is typically\cite{pritchett:2003} $\Omega_{ci}\Delta t < (\Delta x/d_i)^2/\pi$ (where $d_i=c/\omega_{pi}$ is the ion skin depth). In Hybrid-VPIC, as in some other hybrid PIC codes, the field solver may be sub-cycled to use a smaller time step to advance the fields before cycling through the particle advance. This is most useful when the grid resolution is very fine with $\Delta x \ll d_i$.

The hybrid model is closed by providing an evolution equation for the electron pressure $p_e$, either through an equation of state or with an additional time-dependent energy balance equation. For many problems with uniform initial plasma conditions, a simple equation of state of the form
\begin{equation}
    p_e = p_0 \left(\frac{n_e}{n_0}\right)^\gamma
\end{equation}
is sufficiently accurate to capture electron pressure effects. Here, quasi-neutrality is again assumed so that $n_e = \sum_s{q_sn_s}$, and typical choices for the adiabatic index $\gamma$ are $\gamma=1$ (isothermal limit, which works for any system with uniform electron temperature) and $\gamma=5/3$ (adiabatic limit).

For more complicated systems with gradients in the initial conditions, a separate electron energy evolution equation is required. This takes the form:
\begin{equation}
    \frac{\partial p_e}{\partial t} = -\gamma \nabla\cdot (p_e{\bf{u_e}}) + (\gamma-1){\bf{u_e}}\cdot\nabla p_e + (\gamma-1)(-\nabla\cdot {\bf{Q_e}} + H_{ei}),
\label{eq:dpdt}
\end{equation}
where the electron velocity is inferred from ${\bf{u}}_e = -(\nabla\times{\bf{B}}/\mu_0 - {\bf{J}}_i)/en_e$, with ${\bf{J}}_i = n_ee{\bf{\bar{u}}}_i$ the current carried by the ions. If used in the place of a simple equation of state in Hybrid-VPIC, Eq.~\ref{eq:dpdt} is integrated in time within the same numerical loop as the magnetic field evolution. For many problems, the electron heat flux may be modeled with a heat conductivity $\kappa$ as
\begin{equation}
    {\bf{Q_e}} = -\kappa\nabla T_e
\end{equation}
where $\kappa$ = $\kappa_e + \kappa_0$, $\kappa_e$ is given by a physical model and may depend on the local plasma conditions, and $\kappa_0$ is a small ($\kappa_0\sim0.01$---$0.1\kappa_e$ for models with a physical heat conductivity, or $\kappa_0\sim0.01 n_0d_iv_A$ for magnetized simulations) constant numerical diffusion coefficient set separately for convenience that helps maintain numerical stability. Models for electron and ion energy exchange may be included through the term $H_{ei}$, which in Hybrid-VPIC is captured by adding energy lost in each cell by ions in collisions models back to the local electron fluid cell. So far, the separate electron energy equation has primarily been used in Hybrid-VPIC in an electrostatic version for collisional regimes as described in Sec.~\ref{sec:hybrides}.

\subsection{PIC implementation}
\label{subsec:pic}

While the massless electron hybrid PIC equations are relatively standard, there exist a few variations on the numerical methods for their solution. Here, we describe some particular choices in the solver implementation used in Hybrid-VPIC. Hybrid-VPIC uses an explicit time-stepping algorithm, closely resembling the H3D hybrid PIC code \cite{karimabadi:2011}. Although implicit algorithms\cite{stanier:2019} have superior conservation properties, an explicit integrator was chosen for Hybrid-VPIC for its relative simplicity and speed. The leapfrog method is used to advance the particles, so that velocities are known at half time steps:
\begin{equation}
    {\bf{v}}_p^{n+1/2} = {\bf{v}}_p^{n-1/2} + \frac{q_s}{m_s}({\bf{E}}^n + {\bf{v}}_p^n\times{\bf{B}}^n)\Delta t
    \label{eq:vadv}
\end{equation}
where ${\bf{v}}_p^n = 0.5({\bf{v}}_p^{n+1/2} + {\bf{v}}_p^{n-1/2})$ and ${\bf{v}}_p^{n+1/2}$ can be solved for explicitly. We solve this with a standard Boris method\cite{boris:1970}, which accelerates the particle in the electric field over $\Delta t/2$, applies an an energy-conserving rotation about the magnetic field over $\Delta t$, and then accelerates the particle in the electric field over another $\Delta t/2$. The particle positions are known at integral time steps:
\begin{equation}
    {\bf{x}}_p^{n+1} = {\bf{x}}_p^{n} +{\bf{v}}_p^{n+1/2}\Delta t
\end{equation}
The electric field is given by an Ohm's law as in Eq.~\ref{eq:ohm} as a function of the magnetic field and the particle density and current moments, ${\bf{E}}^{n}({\bf{B}}^n,{\bf{J}}_i^n,n_i^n)$. A complication in explicit hybrid PIC codes is that electric field ${\bf{E}}^n$ in Eq.~\ref{eq:vadv} depends itself on the ion current density (a moment of ${\bf{v}}_i^{n+1/2}$ known at half time steps). A few methods for handling this have been developed, including different flavors of the predictor-corrector method \cite{harned:1982,kunz:2014}, the current advance method and cyclic leapfrog\cite{matthews:1994} (CAM-CL) algorithm, and simple linear extrapolation. We use simple linear extrapolation in Hybrid-VPIC because it is efficient (requiring only one particle push per time step) and sufficiently accurate for many problems\cite{karimabadi:2004}. To extrapolate the ion current in Ohm's law, we therefore use:
\begin{equation}
    {\bf{J}}_i^{n+1} =  \frac{3}{2}{\bf{J}}_i^{n+1/2} - \frac{1}{2} {\bf{J}}_i^{n-1/2}
\end{equation}

Unlike the fully kinetic version of the VPIC code, which implements a Yee grid \cite{bowers:2008,yee:1966} for the electromagnetic solver, Hybrid-VPIC uses a cell-centered grid for all field quantities. This includes the electric and magnetic fields as well as the particle moments gathered on the grid. Gradients of a grid quantity $Q$ in the code are computed with standard second-order centered finite differences:
\begin{equation}
    \nabla Q_{i,j,k} = \frac{Q_{i+1,j,k}-Q_{i-1,j,k}}{2\Delta x}{\bf\hat x} +\frac{Q_{i,j+1,k}-Q_{i,j-1,k}}{2\Delta y}{\bf\hat y} +\frac{Q_{i,j,k+1}-Q_{i,j,k-1}}{2\Delta z}{\bf\hat z}
\end{equation}
Each macro-particle's position $\bf{x_p}$ in VPIC is stored as a cell index $I$, which corresponds to a triplet of Cartesian cell indices $I\leftrightarrow (i,j,k)$, along with relative coordinates within the cell $(dx,dy,dz)\in[-1,1]^3$. The particles each belong to a species, which is assigned a charge $q_s=Z_se$ and mass $m_s$. VPIC allows for variable particle weights $w_p$ so that each macro-particle may represent a different number of physical particles. For interpolation to and from the grid, two particle shapes have been implemented in the Hybrid-VPIC code. The first is the simple nearest grid point (NGP) shape. In the NGP scheme, a delta function shape assigns for a particle with a position ${\bf{x}}_p$ in cell $I\leftrightarrow (i,j,k)$ a charge density and current
\begin{equation}  
\begin{cases}
    n_{i,j,k} = q_sw_p/V_c, \\
    {\bf{J}}_{i,j,k} = q_sw_p{\bf{u}}_p/V_c, \\
\end{cases}
\label{eq:ngp}
\end{equation}
with zero contribution at all other cells and where $V_c = \Delta x \Delta y \Delta z$ is the cell volume. For the particle push, the local cell field values at ${\bf{x}}_p^n$ are used. A relatively large number of particles per cell may be required to reduce particle noise to acceptably low levels when using NGP shapes. NGP, however, has the interesting benefit that it gives the only exactly asymptotic-preserving hybrid PIC scheme \cite{stanier:2020} free from spurious numerical dispersion in grids that poorly resolve the ion inertial length. In nonlinear regimes, the numerical dispersion can lead, for example, to an unphysical filamentation of shock fronts (see Sec.~\ref{subsec:explosion}). 

The second particle shape available in Hybrid-VPIC is a quadratic sum (QS) scheme. This is not a standard quadratic B-spline shape used in other PIC codes, which is a tensor product of splines in each direction. The QS scheme was selected to be compatible with the existing parallel communication routines in the fully kinetic version of VPIC. Because the fully kinetic VPIC code uses low-order particle shapes, parallel communication between domains is only necessary for neighbors that share a full face. Each domain in a 3D simulation therefore has only up to 6 neighbors that require communication, rather than the full 26 neighbors that share at least one vertex. In addition to this limitation, VPIC is currently set up with only a single ghost cell at each end of each direction of a domain. The interpolation coefficients for the QS scheme are given in Sec.~\ref{sec:qsappend} of the Appendix. Although the QS does not share the low-order continuity properties of a full tensor product quadratic spline, in practice it gives a similar smoothing of particle noise. In problems that are sensitive to particle noise, it is also possible to apply one or more passes of a self-consistent field and moment smoothing operator in Hybrid-VPIC, following the method of Section 4.3 of \citet{stanier:2019}. This smoothing routine can be applied when using either NGP or QS particle shapes.

The particles are advanced in the electromagnetic fields by re-using the Boris\cite{boris:1970} particle push already implemented in VPIC. The fully kinetic VPIC code is relativistic, and the particle data are stored in terms of the relativistic momentum ${\bf{p}} = \gamma m_s {\bf{u}}$ along with each particle's position. The macro-particles are advanced following the equations of relativistic dynamics. The hybrid PIC model, however, is formulated in a low-frequency limit that does not include light waves, and it is therefore not consistent with a relativistic treatment of the bulk ion motion. The main version of Hybrid-VPIC therefore simplifies the particle push by solving the equations of motion in the non-relativistic limit $\gamma\rightarrow 1$. It is possible, however, to treat plasmas that contain a tenuous population of relativistic ions if the bulk ion flows remain non-relativistic \cite{haggerty:2019}. There is thus the option in Hybrid-VPIC to retain the relativistic particle push.

\subsection{Open boundary conditions}
\label{subsec:openbc}

The simplest boundary conditions in Hybrid-VPIC are fully periodic. In order to expand the range of problems amenable to simulation in Hybrid-VPIC, open boundary conditions were developed that allow plasma and magnetic flux to flow into or out of the simulation domain. Ion particles are absorbed at the open boundaries, and new particles are re-injected if necessary. The injected particle flux is sampled from a drifting multi-Maxwellian velocity distribution using a scheme that matches specified densities, flows, and pressure moments at the boundary \cite{daughton:2006}.

The magnetic field is formally split into two components ${\bf{B}} = {\bf{B_0}} + {\bf{B_1}}$, with a fixed external field $\bf{B_0}$ and a time-varying component $\bf{B_1}$. The external field is a vacuum  field ($\nabla\times{\bf{B_0}}=0$) generated by a system of currents outside the plasma. Examples are the interplanetary magnetic field and planetary dipole field in a global magnetosphere simulation, or the confinement fields generated by external coils in a magnetic mirror device (see example applications in Sec.~\ref{sec:emapplications}). The particles are advanced in the total magnetic field $\bf{B}$. In the field solver, only $\bf{B_1}$ is advanced in time, and $\bf{B_0}$ may be dropped when computing the plasma current density in the Hall term from $\mu_0{\bf{J}} = \nabla\times{\bf{B}} = \nabla\times{\bf{B_1}}$. Note that this is an exact splitting of the $\bf{B}$ field, and not a perturbative approximation based on the size of $\bf{B_1}$. The advantage of splitting $\bf{B}$ in this manner is that for the open boundary conditions, $\bf{B_0}$ is left fixed in the boundary ghost cells. Otherwise, the external field may change over time by diffusing at the boundaries. For the open boundary condition on the time-varying component of $\bf{B}$, the field $\bf{B_1}$ in each ghost cell is set equal to the value of its neighbor within the simulation domain.

The electric field is handled differently. The electric field in ghost cells along an open boundary is advanced in time along with the electric field within the bulk cells. But when computing the electric field from the Ohm's law of Eq.~\ref{eq:ohm} in the a ghost cell, it is assumed that there are no gradients in either $\bf{B}$ or the plasma moments ($n$ and ${\bf{u}}_i$) normal to the boundary. The plasma moments, like the magnetic field, within the ghost cell are set equal to their neighboring values. So, for example, an open boundary in the $x$ direction may have pressure gradient fields $E_y$ and $E_z$ from $y$ and $z$ gradients in $p_e$, but there is no pressure gradient-driven component $E_x$. Crucially, for an open $x$ boundary, gradients in $y$ and $z$ (but not $x$) are retained in the Hall term $\propto (\nabla\times{\bf{B}})\times{\bf{B}}$. 
This treatment of the electric field at open boundaries significantly enhances numerical stability. Nevertheless, it is often necessary to also include a buffer region several cells wide with an enhanced hyper-resistivity $\eta_H$ to dissipate residual oscillations.

\section{Applications of electromagnetic code}
\label{sec:emapplications}

Here, we describe a few applications of the electromagnetic Hybrid-VPIC code to demonstrate the types of problems where the code proves useful. The examples were also chosen to illustrate one or two numerical problems associated with hybrid PIC treatments that are described above in Sec.~\ref{sec:hybridem}.

\subsection{Dayside global magnetosphere}

Hybrid PIC codes have been used for a few decades to simulate the global magnetospheres of magnetized bodies interacting with the solar wind \cite{swift:1995,karimabadi:2006,travnivcek:2007,omidi:2010,lin:2014}. A simple global magnetosphere model has been implemented in Hybrid-VPIC. The model contains a uniform flowing solar wind plasma carrying an interplanetary magnetic field. The magnetic field and flow velocity may be arbitrarily specified in the code. A "planet" is initialized within the domain containing a purely dipole magnetic field. The external field ${\bf{B}}_0$ (see Sec.~\ref{subsec:openbc}) is set equal to the sum of the dipole field and the interplanetary field. Rudimentary inner boundary conditions are set at the planet: the electric field is set to zero within the planet, and particles may be absorbed or reflected off the planetary surface. The open boundaries include the injection of fresh solar wind plasma.

\begin{figure}[h]
\includegraphics[width=120mm]{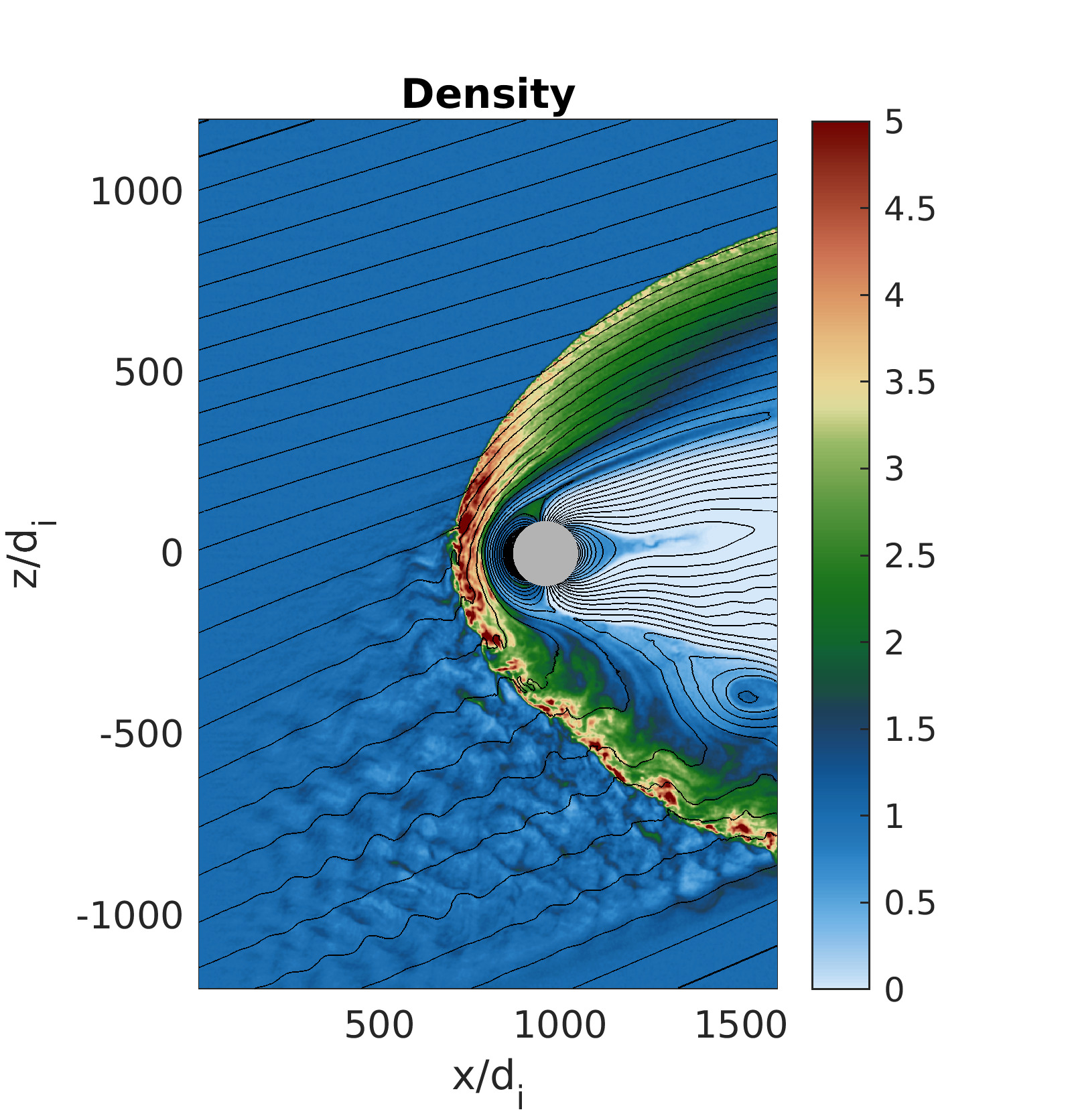}
\caption{\label{fig:dipole} 
Contours of the total plasma density (normalized to the upstream solar wind density) along with sample magnetic field lines in a hybrid simulation of the interaction of the solar wind with a planetary dipole magnetic field. The fluctuations in the lower half are in the foreshock upstream of the quasi-parallel bow shock, and they are driven by ions reflected back upstream from the shock front. 
}
\end{figure}

An example of the plasma density (normalized to the solar wind density) in a 2D Hybrid-VPIC magnetosphere simulation is plotted in Fig.~\ref{fig:dipole}. The solar wind flows from left to right with an Alfven Mach speed of $M_A=10$, and the planet has a radius of $80~d_i$ (where $d_i$ is the ion skin depth in the upstream solar wind). The simulation domain is $L_x\times L_z=1600~d_i\times3200~d_i=800\times1600$ cells, where $d_i$ is the ion inertial length based on the solar wind density. Typical values in the solar wind at 1 A.U.\cite{Klein:2019} are an ion inertial length of $d_i\sim100~km$ and an Alfven speed of $v_A\sim50~km/s$. The electron and ion betas are $\beta_e=\beta_i=0.5$, and the electrons follow an isothermal equation of state. The planetary dipole strength is set so that the magnetopause (the boundary between close magnetospheric field lines and open interplanetary field lines) is $\sim 100~d_i$ above the planet's surface. The time step used is $\Delta t = 0.002/\omega_{ci}$, where $\omega_{ci}$ is the ion cyclotron frequency in the interplanetary magnetic field, and the field solver is sub-cycled 5 times per particle push. Lowest-order NGP particle weights are used with $\sim100$ numerical particles per cell. While this example simulation uses the open boundary conditions described in Sec.~\ref{subsec:openbc}, dissipative layers are placed around the inner planetary boundary and the simulation domain boundaries and additional magnetic field smoothing is used to help with numerical stability (see example input deck \cite{hybridvpic}).

The data in Fig.~\ref{fig:dipole} are plotted at time $t=180/\omega_{ci}$ a little after a full transit time of the solar wind across the simulation domain. The simulation shows the formation of a bow shock along with the magnetosheath, which is the layer of high-density shocked solar wind plasma behind the shock front. The kinetic ions treatment of hybrid PIC codes allows the development of a foreshock, which is a region of turbulence driven by particles reflected back upstream from the shock front. The foreshock fluctuations are strongest at upstream of the quasi-parallel bow shock, where the magnetic field lines are close to parallel to the shock normal. For the orientation of the interplanetary magnetic field in this simulation (the field makes an angle of $20^\circ$ with respect to the inflow $x$ direction), the quasi-parallel shock region is located mainly in the lower half of the simulation domain. In that region, there are order-unity fluctuations in the plasma density. The dominant mode in the foreshock is the electromagnetic resonant ion-ion beam instability \cite{gary:1991,keenan:2022,le:2023} driven by parallel-streaming ions reflected back upstream from the bow shock

\subsection{Astrophysical explosions}
\label{subsec:explosion}

An astrophysical explosion refers to the rapid expansion of high-energy ionized material into a background plasma \cite{winske:2007}, which occurs both naturally and in laser-driven laboratory experiments\cite{clark:2013}. The energy of the debris couples to the background through convective electric fields that pick up the background ions and expel magnetic field, forming a diamagnetic cavity\cite{winske:2019}. In collisionless plasmas, the debris ions may also decouple from the fields if they have large Larmor radii that carry them outside the cavity \cite{hewett:2011,le:2021}, which is one example of why a kinetic treatment of the ions may be important this for problem.

\begin{figure}[h]
\includegraphics[width=170mm]{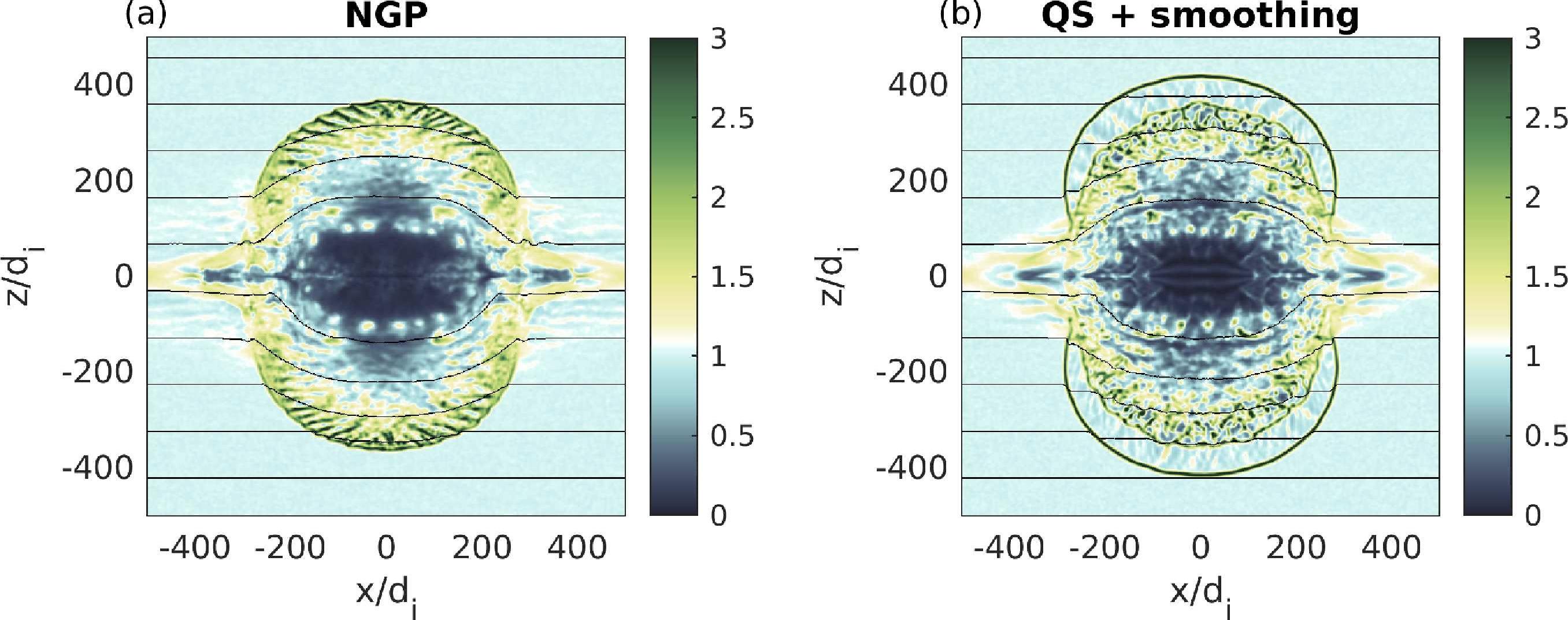}
\caption{\label{fig:explosion} 
Contours of total plasma density (normalized to the background density) from Hybrid-VPIC simulations of an astrophysical explosion--the expansion of a cloud of ionized debris into a background magnetized plasma. The black lines are sample magnetic field lines. These expanding debris initially moves radially outward with an Alfven Mach number of $M_A=15$ and a so-called equal mass radius of $R_m=150~d_i$ (a ball of radius $R_m$ in the background plasma contains the same mass as the exploding  debris ions).  (a) Simulation using nearest grid point (NGP) weights, which does not suffer from numerical dispersion caused by finite particle sizes in hybrid PIC codes\cite{stanier:2020}. (b) Similar simulation, but with quadratic sum (QS) particle weights as described in Sec.~\ref{subsec:pic} and 10 passes of binomial smoothing\cite{stanier:2019} of the moments and fields. The spread of the particle weights over many cells causes an unphysical filamentation of the shock front, with the first filament outrunning the remainder of the shock.  
}
\end{figure}

Figure~\ref{fig:explosion} shows examples of the density profiles in 2D hybrid PIC astrophysical explosion simulations. More examples of Hybrid-VPIC simulations of this type, including a 3D simulation, may be found in previous papers\cite{le:2021,keenan:2022}, along with verification studies that compare the code to linear wave solvers. The simulations initially contain a small Gaussian-shaped cloud of debris ions moving radially outward with an Alfven Mach number (velocity normalized to Alfven speed) of $M_A=15$. The total mass of debris corresponds to an equal mass radius of $R_m=150~d_i$, which means the total debris mass is equal to that of a ball of radius $R_m$ of the background plasma. Both simulations are plotted just a little after the diamagnetic cavity has reached its maximum size and has begun to collapse back down.

The expansion of the debris drives an outward propagating shock in the background plasma. These simulations have domain sizes of $L_x\times L_z=1000~d_i\times1000~d_i$, and they are intentionally somewhat under-resolved with 512 cells in each direction. A time step of $\Delta t=0.01/\omega_{ci}$ is used, and there are 200 numerical particles per cell in the background and $\sim1$ million numerical particles representing the debris cloud. The background plasma beta is $\beta_i=\beta_e=0.1$, with the electron obeying an isothtermal equation of state. The input deck for this example is available online \cite{hybridvpic}.

The coarse resolution of the grid helps highlight a numerical problem mentioned in Sec.~\ref{subsec:pic} that is caused by spreading particle weights over several cells in poorly-resolved hybrid PIC simulations\cite{stanier:2020}. Although problem-dependent, a grid resolution of $\Delta x/d_i\leq 1$ is recommended to avoid this numerical issue. The NGP weighting used in Fig.~\ref{fig:explosion}(a) does not display this problem, and the shockfront retains a physically realistic structure. In Fig.~\ref{fig:explosion}(b) with a QS particle weight scheme and 10 passes of particle moment and field smoothing routine, however, the shockfront breaks apart into filaments. The initial filament at the leading edge of the shock also runs away from the rest of the shock at an unphysically high speed. Therefore, while higher-order particle shapes or smoothing can help reduce the effects of particle noise, they can also induce spurious unphysical dispersion. The best choice of particle shape and number of smoothing passes depends on the particular problem and the grid resolution afforded by the available computing resources.

\subsection{Magnetic mirror fusion device}
As a final example application, we describe a Hybrid-VPIC model of a magnetic mirror fusion experiment. Magnetic mirrors have received renewed interest recently because of the availability of high-field superconducting coils and promising results from the Gas Dynamic Trap (GDT) experiment with confinement approaching theoretical limits \cite{ivanov:2003,ivanov:2013}. A kinetic ion model of the GDT concept is necessary for a few reasons. First, all mirror machines include loss-cone velocity distributions that may seed kinetic instabilities. Also, while the GDT includes a relatively collisional thermal plasma in the central core, the device is then fueled with multi-MeV ions injected as initially neutral beams. The fast fuel ions are nearly collisionless, and the fusion performance is sensitive to the slowing down and velocity distribution of the fast ions. Last, finite Larmor radius (FLR) effects are expected to stabilize MHD interchange modes, one of the main instabilities that degrade mirror confinement.

A mirror machine model was implemented in Hybrid-VPIC by first initializing the magnetic field ${\bf{B}}_0$ based on a nominal external coil configuration. Any coil configuration may be specified by the user.  The boundaries are open, and they are set to cut out any region where the external coils intersect the Cartesian volume of the domain. A thermal plasma is initialized in the central region of low magnetic field. Because particles in the loss cone escape from the plasma, a volumetric source in included in the center of the mirror to replenish the plasma density. We use a source with a Maxwellian velocity distribution and with a total flux of new particles set to maintain a steady running average of the central plasma density. The particle source term may also be arbitrarily set by the user. To include fast ions injected as neutral beams, an additional volumetric source may be included with ions sampled from a beam-like distribution, peaked at experimentally relevant pitch angles near $45^\circ$.  

\begin{figure}[h]
\includegraphics[width=170mm]{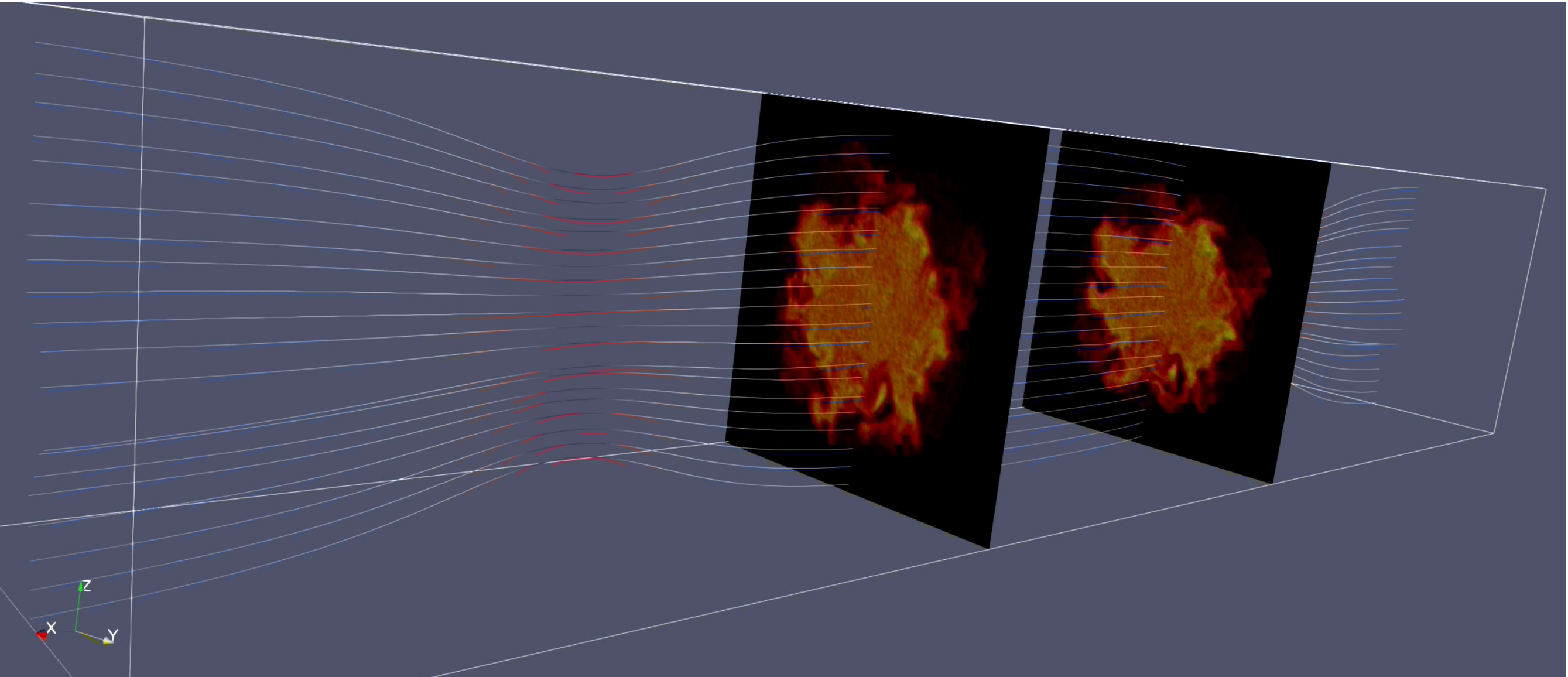}
\caption{\label{fig:mirror} 
Two slices of the plasma density from a Hybrid-VPIC simulation of a magnetic mirror confinement device are shown along with sample magnetic field lines. This set-up has a mirror ratio of $B_m/B_0=5$ and a central beta of $\beta_i=\beta_e=0.03$. The 3D simulation shows the development of interchange modes at the plasma edge. Finite Larmor radius (FLR) effects should stabilize these MHD modes, and this problem is amenable to study with a hybrid PIC code. 
}
\end{figure}

Sample data from a 3D Hybrid-VPIC simulation of a mirror are plotted in Fig.~\ref{fig:mirror}. The simulation domain is $L_x\times L_y\times L_z = 300~d_i \times 30~d_i\times30~d_i = 800\times160\times160$ cells, and it uses 50 particles per cell at the peak of the density profile. An external magnetic field coil geometry is supplied that results in a mirror ratio (peak magnetic field on axis normalized to central magnetic field) of 5. The electron and ion temperatures are equal and result in plasma betas of $\beta_i=\beta_e=0.03$. The electrons follow a simple isothermal equation of state, though modifications can be made to account for more realistic electron profiles\cite{wetherton:2021} in mirror devices. The run has a time step of $\Omega_{ci}\Delta t = 0.01$ (with $\Omega_{ci}$ evaluated at the peak on-axis magnetic field location), and it reaches a quasi-steady state within $\sim 150,000$ time steps. The simulation ran in $\sim12$ hours on 8 nodes of a computing cluster with 128 CPU cores per node. Apparent in Fig.~\ref{fig:mirror} are interchange modes at the edge of the plasma density column. These MHD modes are driven by the plasma pressure gradient in the presence of magnetic curvature. Interchange modes are one of the main loss mechanisms in mirror devices. The shorter wavelength modes, however, are believed to be stabilized by ion FLR effects. We plan to model this in more detail in the future with Hybrid-VPIC.

\section{Electrostatic Collisional Hybrid Model}
\label{sec:hybrides} 

One of the main applications of the Hybrid-VPIC code has been studying kinetic mix in high-energy density (HED) plasma regimes relevant to inertial confinement fusion (ICF). Hybrid PIC models have previously been able to demonstrate such kinetic mix or species separation effects in 1D simulations \cite{bellei:2013,le:2016,sio:2019}. The high performance of Hybrid-VPIC has allowed this to be extended into 2D simulations that include turbulent mixing, and example results are described in more detail in Sec.~\ref{sec:interface}. Hybrid-VPIC also includes for the first time in a hybrid PIC code an improved electron fluid transport model for multi-ion species collisional plasmas\cite{simakov:2014}. These studies have so far focused on the simpler electrostatic regime without an externally imposed or self-generated\cite{walsh:2017,sadler:2020} magnetic field. In this case the magnetic terms are dropped from the Ohm's law, and the electric field takes the form:
\begin{equation}
{\bf{E}} = - \frac{1}{en_e}\nabla p_e + {\bf{R}}_{ei}
\label{eq:esohm}
\end{equation}

While the electrostatic field model is relatively simple, a more detailed treatment is necessary for Coulomb collisions in ICF-relevant HED regimes. For ion-ion collisions, we re-use the Takizuka-Abe \cite{takizuka:1977} particle-pairing routine that has been used extensively in the fully kinetic VPIC code \cite{daughton:2009,roytershteyn:2010,le:2015}. We note that a  similar binary pairing algorithm for fusion burn has also been implemented in Hybrid-VPIC\cite{higginson:2019,le:2016}, though this is not a necessary component for the mix modeling described in Sec.~\ref{sec:interface}.

Ion-electron collisions are treated primarily through a particle-fluid collision model (which we call $C_{ie0}\{f_s\}$ for each ion species $s$) based on collision frequencies of ions colliding with a Maxwellian electron distribution. An additional correction term is added to give the full ion collision operator $C_i\{f_s\} = C_{ie0}\{f_s\} + C_{th}\{f_s\} $, where $C_{th}\{f_s\} $ accounts for the thermal force. The main ion-electron collision operator $C_{ie0}\{f_s\}$ is handled with a Monte-Carlo model for particle scattering off a fluid background \cite{lemons:2009}. One advantage of the particle-fluid scattering model over a reduced fluid model is that it allows a better treatment of tenuous populations of fast ions, such as alphas produced by fusion, for which the assumption of $v_i\ll v_{the}$ of typical fluid expansions is not well-satisfied. Using the local electron density and temperature within each cell, the appropriate pitch angle, friction, and energy diffusion scattering rates for a given ion are computed. From these, a frictional drag term and a pseudo-random scattering angle are computed.

As mentioned above, an additional term is included in the scattering operator to account for the thermal force, which depends on low-order non-Maxwellian corrections to the electron distribution in the presence of temperature gradients. To compute the thermal force, we use the transport coefficients for a multi-ion collisional plasma calculated by \citet{simakov:2014}. The thermal force collision operator $C_{sth}\{f_s\}$ gives a additional acceleration of each ion macro-particle. The additional acceleration of each ion $p$ is:
\begin{equation}
    {\bf{a}}_{th} = \frac{\beta_0 Z_p^2}{m_p \sum_s(n_sZ_s^2)} \nabla T_e,
    \label{eq:fthermal}
\end{equation}
where the transport coefficient $\beta_0(Z_{eff})$ is given\cite{simakov:2014} by a Pad\'e approximant in terms of of $\Zeff = \sum_sn_sZ_s^2/n_e$. Because the model depends on $\Zeff$, during the main particle loop of the PIC code, we collect the $(Z_s)^2$-weighted moments of the particles. Explicitly, we gather ${\bf{J}}_{2i} = \sum_s n_s Z_s^2 {\bf{u_s}}$ and $n_{2i} = \sum_s n_s Z_s^2 $ at each grid point in addition to the standard $(Z_s)$-weighted current and charge densities.

The scattering operator is applied with a lowest-order operator splitting method, whereby the collision velocity update of each ion is performed after the Boris push in the electric field. $C_{ei0}\{f_s\}$ uses the velocity transformation of \citet{lemons:2009}, and $C_{th}\{f_s\}$ simply adds ${\bf{a}}_{th}\Delta t$ to each ion's velocity. To conserve total energy and momentum, the momentum increment $\Delta (m_p {\bf{u_p}})$ of each ion macro-particle is added back into the Ohm's law for the electric field (giving the resistive term ${\bf{R}}_{ei}$ in Eq.~\ref{eq:esohm}) and the energy increment $\Delta (m_p u_p^2/2)$ is included in the term $H_{ei}$ of Eq.~\ref{eq:dpdt} as a local heat source for the electrons.

The electron energy balance equation of Eq.~\ref{eq:dpdt} is solved in the electrostatic approximation ${\bf{u}}_e = \sum_s n_sZ_s{\bf{u}}_s/\sum_s n_sZ_s$ using the electron heat flux model for ${\bf{Q}}_e$ from Eq. (13) of \citet{simakov:2014}:
\begin{equation}
    {\bf{Q}}_e = \beta_0 p_e ({\bf{u}}_e - {\bf{\hat u}}_i) - \kappa\nabla T_e
\end{equation}
where ${\bf{\hat u}}_i = \sum_s n_sZ_s^2 {\bf{u_s}}/ \sum_s n_sZ_s^2$, $\kappa = \kappa_e + \kappa_0 = \gamma_0 p_e/m_e\sum_i\nu_{ei} + \kappa_0$, $\kappa_0$ is a small ($\kappa_0<\kappa_e$ under nominal plasma conditions) numerical diffusion constant as in Sec.~\ref{sec:hybridem}, $\nu_{es} = \sqrt{32\pi}n_sZ_s^2e^4 \log\Lambda/3m_e^{1/2}T_e^{3/2}$, $\beta_0$ is the same coefficient as in Eq.~\ref{eq:fthermal}, and the transport coefficient $\gamma_0$ is given\cite{simakov:2014} by an approximation of another function of $\Zeff$. Note that while the electron viscous transport coefficients in this limit are also known\cite{simakov:2014}, the electron viscosity has not yet been implemented in Hybrid-VPIC because it is usually less important than ion viscosity.

\section{Application: Plasma kinetic effects on interfacial mix }
\label{sec:interface}

\begin{figure*}
\includegraphics[width=166mm]{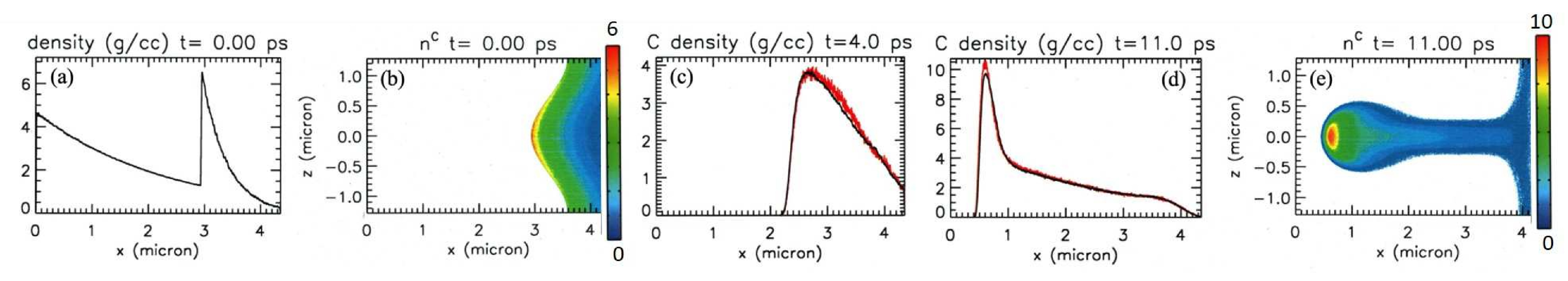}
\caption{\label{fig:1} 
Hybrid simulation of  Rayleigh-Taylor instability growth of a D-C interface
in the presence of a steep density gradient with $g=2.1 \times 10^{19}$~cm/s$^2$:
(a) Initial density (in g/cc) profile at $z=0$ with isothermal hydrostatic equilibrium, (b) C ion density at $t=0$ showing the initial D-C interface configuration of the single-mode perturbation with $w=2.56$~\microns and $a=0.64$~\microns ($a/w=0.25$), (c)-(d) C ion density profile at $z=0$ at time $t=4.0$~ps and 11~ps, comparing results between the hybrid (black curves) and the fully-kinetic VPIC code (red curves), (e) C ion density at $t=11.0$~ps from the hybrid simulation. Gravity is in the $-\hat{x}$ direction.}
\end{figure*}

In this section, we present in some detail an application of Hybrid-VPIC to model the plasma kinetic effects on interfacial mix in HED settings 
involving ambipolar diffusion, viscosity, species mass transport~\cite{molvig:2014,molvig:2014pop,haines:2014,simakov:2016,simakov:2016b,vold:2017,vold:2018,vold:2018b,yin:2016,yin:2019}, and hydrodynamic instabilities.
We begin with the modeling of the Rayleigh-Taylor (RT) instability
by imposing a uniform effective gravity with constant $g$ (added as an acceleration in addition to that of the  electric field in the particle push)
on a deuterium-carbon (D-C) interface initialized in hydrostatic equilibrium.
We consider two possible hydrostatic equilibria, isothermal and ``isochoric'' conditions 
(constant density in each material region).

Under isothermal conditions with $T_i = T_e = T$ (constant over the simulation domain), 
an initial density gradient in $\unitvec{x}$ is required to maintain pressure balance across the interface
in the presence of the gravitational force in the $-\unitvec{x}$ direction~\cite{yin:2019}.
Let the species $i$ and $j$ be the light- and heavy-ion species situated across the interface in $\unitvec{x}$.
Summing the ion and electron momentum equations
\bearr
m_i n_i {d \V{v}_i \over dt} = - T \nabla n_i - m_i n_i g \unitvec{x}
\\
m_e n_e {d \V{v}_e \over dt} = - T Z_i \nabla n_i - m_e Z_i n_i g \unitvec{x}
,
\eearr
setting the convective derivatives equal to zero, and taking $m_i \gg m_e$, 
one obtains for species $i$,
$\partial  \log n_i / \partial x = m_i g / \left[ (Z_i+1) T \right]$. 
Integrating this yields the density profile
\be
n_i(x) = n_{i0} \exp\left[- {{m_i g x} \over T (Z_i+1)} \right]
\ee
where $n_{i0}$ is the ion density of species $i$ at $x=0$. 
To maintain pressure balance in $x$ across an interface located at position $x_s(z)$,
the ion density of species $j$ at the interface is required to satisfy
\be
n_j(x,z)  = n_{j0} \exp\left[- {{m_j g [x - x_s(z)]} \over T (Z_j+1)} \right]
\ee
where
\be
n_{j0}(z)  = n_{i0} {Z_i + 1 \over Z_j + 1}\exp\left[- {{m_i g x_s(z)} \over T (Z_i+1)} \right]
.
\ee

Alternatively, we can consider the case where the initial density is constant in a region and 
the temperature $T_e(x) = T_i(x) = T(x)$ is varied to establish hydrostatic equilibrium. 
Summing the
electron and ion momentum equations and taking the limit $d\V{v}_e/dt = d\V{v}_i/dt \rightarrow 0$, we obtain  
\be
\nabla( n_e T_e + n_i T_i ) = - (n_e m_e + n_i m_i) g \unitvec{x}
,
\ee
which implies that 
\be
(Z_i+1) \nabla T = - m_i g \unitvec{x}
,
\ee
or, integrating, 
\be
T(x,z) \approx T_0 - {m_i g \over Z_i + 1} \left[x - x_s(z)\right]
,
\ee
\be
T(x,z) \approx T_0 - {m_j g \over Z_j + 1} \left[x - x_s(z)\right]
,
\ee
where $T_0$ is an integration constant equal to the the temperature at the interface. 
The condition for the pressure gradient to vanish across the material interface
is that the density of the two species satisfies 
\be
n_j = n_i {Z_i+1 \over Z_j + 1}
.
\ee

The hybrid RT simulation is initialized according to the above density or temperature profiles
using a step function plasma interface with a sinusoidal perturbation.
The classical Rayleigh-Taylor growth rate is
$\gamma = \sqrt{A g k}$, 
where $A= (\rho_h - \rho_l)/(\rho_h + \rho_l)$  
is the Atwood number computed from the light- and heavy-ion densities $\rho_l$ and $\rho_h$ 
and $k = 2 \pi / w$ is the wavenumber of the perturbation with wavelength $w$.
However, in the presence of plasma kinetic effects (finite diffusivity and viscosity)
which inhibit the growth at small wavelengths, 
the growth rate is a non-monotonic function of wavenumber $k$ with a maximum occurring at a low $k$ value,
as discussed in Refs.~\cite{yin:2019,vold:2021}, 
 behavior that differs from the classical growth rate expression. 

Hybrid simulations benchmarked with the fully-kinetic VPIC code~\cite{bowers:2008,bowers:2008b,bowers:2009,bird:2021}
of the RT dynamics of a D$^+$-C$^{6+}$ interface 
satisfying hydrostatic equilibrium are discussed here.  
In the simulations,  
the D and C ion number densities at the interface location
are $n_{\rm D}= 5.40 \times 10^{23}$~cm$^{-3}$
and $n_{\rm C}= 1.54 \times 10^{23}$~cm$^{-3}$ 
(the electron number density on the C side is $n_e = 9.24 \times 10^{23}$~cm$^{-3}$)
and $A=0.67$.
The simulations are performed in 2D in the $(x, z)$ plane
with interface position expressed as 
$x_{\mathrm{it}}(z) = L_x/2 - a \cos(2 \pi z / w)/2$, 
where
$L_x $ is the size of the simulation domain in $x$,
$w$ is the wavelength of the single, sinusoidal perturbation along $z$, 
and $a $ is the perturbation peak-to-peak amplitude.
The size of the simulation domain in the $z$ direction is 
set to be equal to the wavelength of the perturbation $w$.

In the simulations, 
we apply a binary collision model~\cite{takizuka:1977} with a Coulomb logarithm $\log \Lambda$ for all collisions.
To conserve energy and momentum rigorously in the collision operator,
all computational macro-particles have the same statistical
weights, i.e., each particle represents the same number of physical particles in a given simulation.
The number of particles per cell is a function of the density.
At the interface location, we employ 167 and 584 computational macro-particles per cell for C and D ions, respectively. 

The collision operator may be applied every $N$ time steps 
with condition $2 dt_{\rm coll} \nu_{\rm ij} \ll 1$
to ensure accuracy of collision sampling~\cite{lemons:2009},
where $dt_{\rm coll} = N dt$ is the sub-cycling time step of the collision operator,
$dt$ is the simulation time step, 
$\nu_{\rm ij}$ is the collision rate for species $i$ and $j$ 
(including self-collisions for $i=j$ and cross-species collisions for $i \neq j$).
In the fully kinetic VPIC simulations, 
the cell size is on the order of the Debye length at the interface,
the time step required by the Courant condition is much smaller than that used in the hybrid code,
and $N > 1$. 
In the hybrid simulations, we use $N=1$ and a time step $dt$ to
obtain the same values of $2 dt_{\rm coll} \nu_{\rm ij} $
as those in the full VPIC simulations. The electron pressure evolution equation is sub-cycled to remain below the CFL condition imposed by the electron heat flux diffusion equation.
The hybrid simulations are run up to $22 \times 10^6$ time steps 
and energy is conserved to within $< 2.5$~\%.

The simulations performed in the $(x, z)$ plane
use periodic boundary conditions on particles and fields in $z$.
The field boundary conditions in $x$ are such that the boundaries are perfect electrical conductors. 
To avoid boundary effects in the simulations, 
the simulation length is long in $x$ and the boundaries are sufficiently far from the interface
that the electric fields are effectively zero at the $x$-boundaries.
``Maxwellian reflux'' boundary conditions are used in $x$, 
where particles encountering a boundary are reinjected from a Maxwellian at the initial local temperature.

The first test problem for the hybrid code, which is benchmarked with the fully-kinetic VPIC code, 
is the RT dynamics of a D-C interface
satisfying isothermal hydrostatic equilibrium initially 
in the presence of a steep density gradient with $g=2.1 \times 10^{19}$~cm/s$^2$.
To reduce the the computational cost for the fully-kinetic VPIC code,
the atomic mass number for C is artificially increased to $A_C = 36$ to enhance the growth rate.
The simulation domain has 
$L_x = 4.35$~\microns 
$L_z=w=2.56$~\micron,
and a perturbation amplitude $a=0.64$~\microns ($a/w=0.25$). 
The wavelength is chosen to be around the maximum growth found under these parameter settings~\cite{yin:2019}.
The temperatures are $T_e = T_i = 5$~keV throughout the simulation domain.

The plasma kinetic effects depend on the collisionality.
In the limit of  high collisionality,  plasma kinetic effects become insignificant. 
On the other hand, if the kinetic effects dominate as in the case for short wavelength modes, 
the effective surface tension can flatten the interface perturbation.
Thus in the simulations, 
we apply the collision model with an enhanced Coulomb logarithm $\log \Lambda = 40 $
in order to control the plasma kinetic effects 
for the choice of our plasma density and perturbation wavelength.
(One could use a normal value of Coulomb logarithm 
and a higher density to obtain the desired collision rates;
however, it is more computationally expensive for the fully kinetic simulations 
since modeling higher electron density plasma media with smaller Debye lengths
would require smaller cell sizes and time-steps.)

Results from RT simulations using an initial isothermal hydrostatic equilibrium 
and a steep density gradient with $g=2.1 \times 10^{19}$~cm/s$^2$
are shown in Fig.~\ref{fig:1}.
The steep density profile along $x$ and 
the interface structure are indicated in frames (a) and (b).
 Gravity is in the $-\hat{x}$ direction and the heavy fluid (C ions) is on right of the light fluid (D ions).
The imposed mode amplitude grows to a nonlinear stage as
the heavy fluid moves into the light fluid, as shown in frame (e).
Plasma kinetic effects impede the growth of short wavelength modes
and small-scale structure is absent as in the full VPIC simulation. 
In contrast, in inviscid hydrodynamics, Kelvin-Helmholtz
instabilities often appear at the edges of an RT mode structure.
The C ion density profiles at $z=0$ at time $t=4.0$~ps and 11~ps from the hybrid simulation
are overlaid with those from the full VPIC code in frames (c) and (d),
showing good agreement.
Note that the cell sizes in the hybrid simulations are 5-6 times larger 
than those in the full VPIC simulation
and the time step is 26 times larger.
Thus, the hybrid simulation not only captures the essential RT dynamics
but also at  a significant cost reduction compared with the full VPIC 
by nearly 3 orders of magnitude.

\begin{figure*}
\includegraphics[width=166mm]{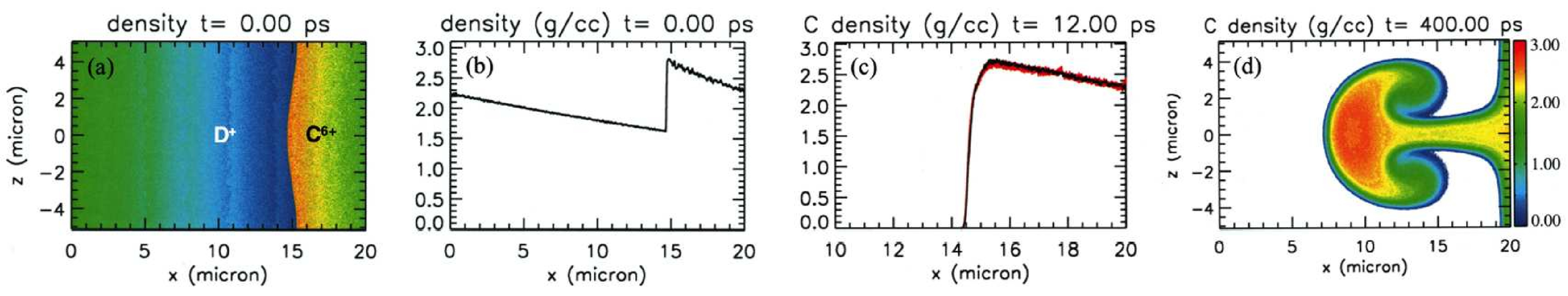}
\caption{\label{fig:2} 
Hybrid simulation of  Rayleigh-Taylor instability growth of a D-C interface at a longer time scale that is 
relevant to ICF settings 
with $g=2.1 \times 10^{17}$~cm/s$^2$
and convergence verification on grid resolution: 
(a) Initial density 
showing the D-C interface configuration of the single-mode perturbation
with $w=10.24$~\microns and $a=0.64$~\microns ($a/w=0.0625$),
(b) initial density profile at $z=0$ with the isothermal hydrostatic equilibrium,
(c) C ion density profile at $z=0$ at time $t=12.0$~ps
comparing results from high-resolution (red curve) and reduced-resolution simulation (black curve),
(d) C ion density at long-time scale $t=400.0$~ps
from the reduced-resolution hybrid simulation.
Gravity is in the $-\hat{x}$ direction.
}
\end{figure*}

Next, we discuss 
hybrid simulations of the RT instability  
at a longer time scale
for $g=2.1 \times 10^{17}$~cm/s$^2$ using the isothermal hydrostatic equilibrium.
The initial D and C ion number densities at the interface location
and the Atwood number are the same as the case above
but without the use of artificial atomic mass number or Coulomb logarithm.
The simulation has 
$T_e = T_i = 1$~keV,
Coulomb logarithm $\log \Lambda = 5$,
$L_x = 20.0$~\micron, 
$L_z = w= 10.24$~\micron,
and $a=0.64$~\microns ($a/w=0.0625$).
These parameters are relevant to the conditions during the deceleration phase of 
a carbon shell imploding onto deuterium fuel in an ICF setting,
where the maximum growth is found to be at a perturbation wavelength of 10~\micron~\cite{vold:2021}.
To examine convergence with grid resolution, two hybrid simulations were run: 
the first, with high spatial resolution: 
$n_x=19040$ and $n_z=9200$,
comparable to those used for the results in Fig.~\ref{fig:1}
where agreement with full VPIC simulation is obtained; the second, with reduced spatial resolution, 
$n_x=1280$ and $n_z=640$
but run to longer times (half nanosecond).

The results are given in Fig.~\ref{fig:2}
with frames (a) and (b) showing the initial conditions.
In frame (c), the C ion density profiles at $z=0$ at time $t=12.0$~ps
from high-resolution (red curve) and reduced-resolution  (black curve) simulations 
are overlaid, showing good agreement.
With convergence verified, the late time RT dynamics from the reduced-resolution simulation
is presented in frame (d)
where small-scale structures are absent 
due to plasma kinetic effects impeding the growth of short wavelength modes.
The behavior is consistent with that from the \textsc{xRAGE} \cite{gittings:2008,haines:2017} simulation
performed with a plasma transport model in the limit of fluid approximations~\cite{vold:2021}.
(We will come back to discuss the implications of this results after presenting multi-mode RT simulations below
for deceleration phase of ICF implosion.)
The reduced-resolution simulation completed over 24 million time steps with energy conservation
within 0.2\%. 

\begin{figure*}
\includegraphics[width=166mm]{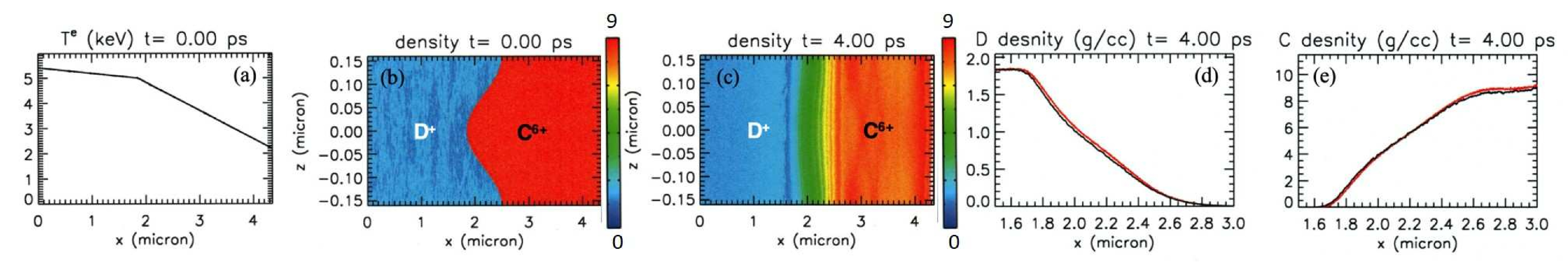}
\caption{\label{fig:3} 
Hybrid simulation of  Rayleigh-Taylor instability growth 
with the isochoric hydrostatic equilibrium conditions and with $g=2.1 \times 10^{18}$~cm/s$^2$: 
(a) Initial temperature profile at $z=0$,
(b) density at $t=0$ showing the initial D-C interface configuration of the single, short-wavelength perturbation
with $w=0.32$~\microns and $a=0.64$~\microns ($a/w=2$),
(c) density at $t=4.0$~ps from the hybrid simulation, 
showing the flattening of the interface from strong kinetic effects on the short-wavelength mode,
(d)-(e) D and C ion density profiles at $z=0$ at time $t=4.0$~ps,
comparing results between the hybrid (black curves) and the fully-kinetic VPIC code (red curves).
Gravity is in the $-\hat{x}$ direction.
}
\end{figure*}

Now we discuss the simulations of RT instability growth 
with the isochoric hydrostatic equilibrium.
The simulation parameters are similar to those used for Fig.~\ref{fig:1}
but with $g=2.1 \times 10^{18}$~cm/s$^2$
and an enhanced Coulomb logarithm $\log \Lambda = 100 $.
The simulation domain has 
$L_x = 4.35$~\microns 
$L_z=w=0.32$~\micron,
and a perturbation amplitude $a=0.64$~\microns ($a/w=2$). 
The short wavelength is chosen to show strong kinetic effects on the interface structure~\cite{yin:2019}.
The temperatures are $T_e = T_i = 5$~keV at the interface location.

The results are shown in Fig.~\ref{fig:3}
where frames (a) and (b) show 
the initial temperature profile at $z=0$
and the initial interface configuration of the short-wavelength perturbation.
Due to the short-wavelength mode, 
the kinetic effects are strong, 
 impeding the growth of the RT mode and flattening the interface on a short time scale,
similar to the dynamics found in full VPIC simulations ~\cite{yin:2019}.
Furthermore, comparisons between the hybrid (black curves) and the fully-kinetic VPIC code (red curves)
for the D and C ion density profiles at $z=0$ at time $t=4.0$~ps
are provided in frame (d), evidencing good agreement.

\begin{figure*}
\includegraphics[width=166mm]{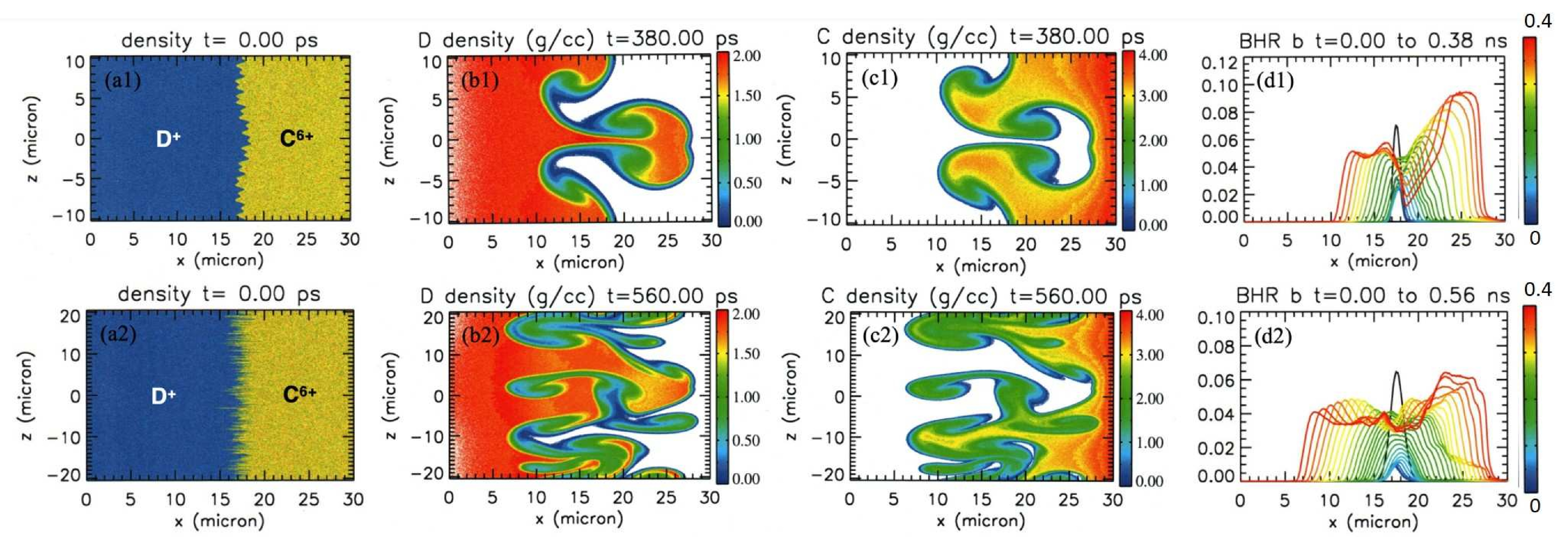}
\caption{\label{fig:4} 
Hybrid simulations of multi-mode Rayleigh-Taylor instability growth of a D-C interface at a longer time scale that is 
relevant to ICF settings 
with $g=2.1 \times 10^{17}$~cm/s$^2$ and the isochoric hydrostatic equilibrium conditions: 
(a1) and (a2) Initial density 
showing the D-C interface configuration of the 4-mode 
and 209-mode simulations, respectively.
(b1) and (b2) D ion density at time $t=380.0$~ps and 560~ps
from the the 4-mode and 209-mode simulations, respectively;
similarly, (c1) and (c2) are for the C ion density.
(d1) and (d2) Evolution for the BHR b parameters vs. time
(early- to late-time profiles are indicated by the spectrum of colors from back to red).
Gravity is in the $-\hat{x}$ direction.
}
\end{figure*}

We have also considered the long time-scale dynamics of multi-mode Rayleigh-Taylor instability growth 
relevant to ICF settings with $g=2.1 \times 10^{17}$~cm/s$^2$ using the isochoric hydrostatic equilibrium conditions.
Two hybrid simulations with 4 modes and 209 modes are performed
using the same D-C interface plasma density and temperature as those for Fig.~\ref{fig:2}
The 4-mode simulation has  
$L_x = 30.0$~\micron, 
$L_z = 20.48$~\micron,
wavelengths $L_z, L_z/2, L_z/10, L_z/22$ for sinusoidal perturbations with phase shifts
and amplitude $a=0.64$~\micron.
The mode with wavelength $L_z/2=10.24$~\microns is the maximum growth mode modeled in Fig.\ref{fig:2}
using the isothermal hydrostatic equilibrium.
The 209-mode simulation has  
$L_x = 30.0$~\micron, 
$L_z = 40.96$~\micron,
wavelengths $L_z/N$, where $N$ is the mode number ranging from 1 to 209,
for sinusoidal perturbations with random phase shifts
and amplitude $a=0.16$~\micron.
The mode with wavelength 10.24~\microns at the maximum growth is also included.

The results from the multi-mode hybrid simulations of RT are shown in Fig.~\ref{fig:4}
with the upper frames for the 4-mode simulation
and the lower frames for the 209-mode simulation.
The initial interface configurations are in frames (a1) and (a2). 
The initial perturbation amplitude is 4$\times$  smaller in the 209-mode simulation
so it took  longer for the modes to evolve.
The energy conservation is $\sim 2.5$\% for these multi-mode hybrid simulations. 
In these multi-mode simulations, the RT dynamics are modeled for a long time-scale ($\sim$0.5~ns)
but the turbulence state is absent.
In contrast, the mode with wavelength 10.24~\microns at the maximum growth
dominates the late-time material field,
as evident in results shown in frames (b1), (b2), (c1), and (c2).

The degree of mixing can be evaluated using the specific-volume-density covariance quantity
\be
b = \left<\rho \right> \left<{1 \over \rho} \right> - 1
\ee
from the Besnard-Harlow-Rauenzahn (BHR) turbulent mix model~\cite{besnard:1992,banerjee:2010}. By construction,
$b$ is nonnegative and in variable density flows such as these where mean pressure gradients are important, $b$  
plays a central hydrodynamical role, describing the material field as well as 
affecting the generation and evolution of turbulence. The quantity $b$ also encodes information about 
the morphology (i.e., the degree of ``mixedness'') of a multi-component plasma and in media
undergoing binary reactions such as thermonuclear reactions, $b$ can be used to modify the 
reaction rates between species in a manner that is faithful to the underlying morphology~\cite{ristorcelli:2017}. Specifically, as $b$ approaches zero, the medium becomes fully atomistically mixed; 
regions of finite $b$ correspond to incomplete mixing. Reaction rate modifications associated 
with such variations in $b$ have been observed in recent separated reactants experiments at the National Ignition Facility~\cite{albright:2022}. 

In Fig.~\ref{fig:4} (d1) and (d2), 
the evolution of BHR $b$ vs. time as a function of $x$, averaged over $z$, 
is shown from the two multi-mode hybrid simulations
(early- to late-time profiles are indicated by the spectrum of colors from back to red).
During the duration of the simulations, 
the quantity $b$ is observed to first decrease and then increase with time as the width of finite $b$ broadens.
From the simulations, we see that this corresponds to a layer of incomplete mixing growing from the 
interface as the largest wavelength modes evolve to finite amplitude. 
The incomplete mix layer width grows and extends over a large fraction of the simulation volume. 
As the simulation time scales are comparable to those of the deceleration phase 
for a carbon shell imploding onto a deuterium fuel for ICF experiments, 
this suggests insufficient time for turbulence to develop and for complete turbulent mixing to occur, 
consistent with findings from prior studies~\cite{robey:2003,weber:2014,weber:2015}. To build on the success of the recent demonstration of ICF ignition \cite{abu:2022}, this hybrid capability can be used to gain understanding of the kinetic effects on mix and burn propagation\cite{daughton:2023}.

\begin{figure*}
\includegraphics[width=166mm]{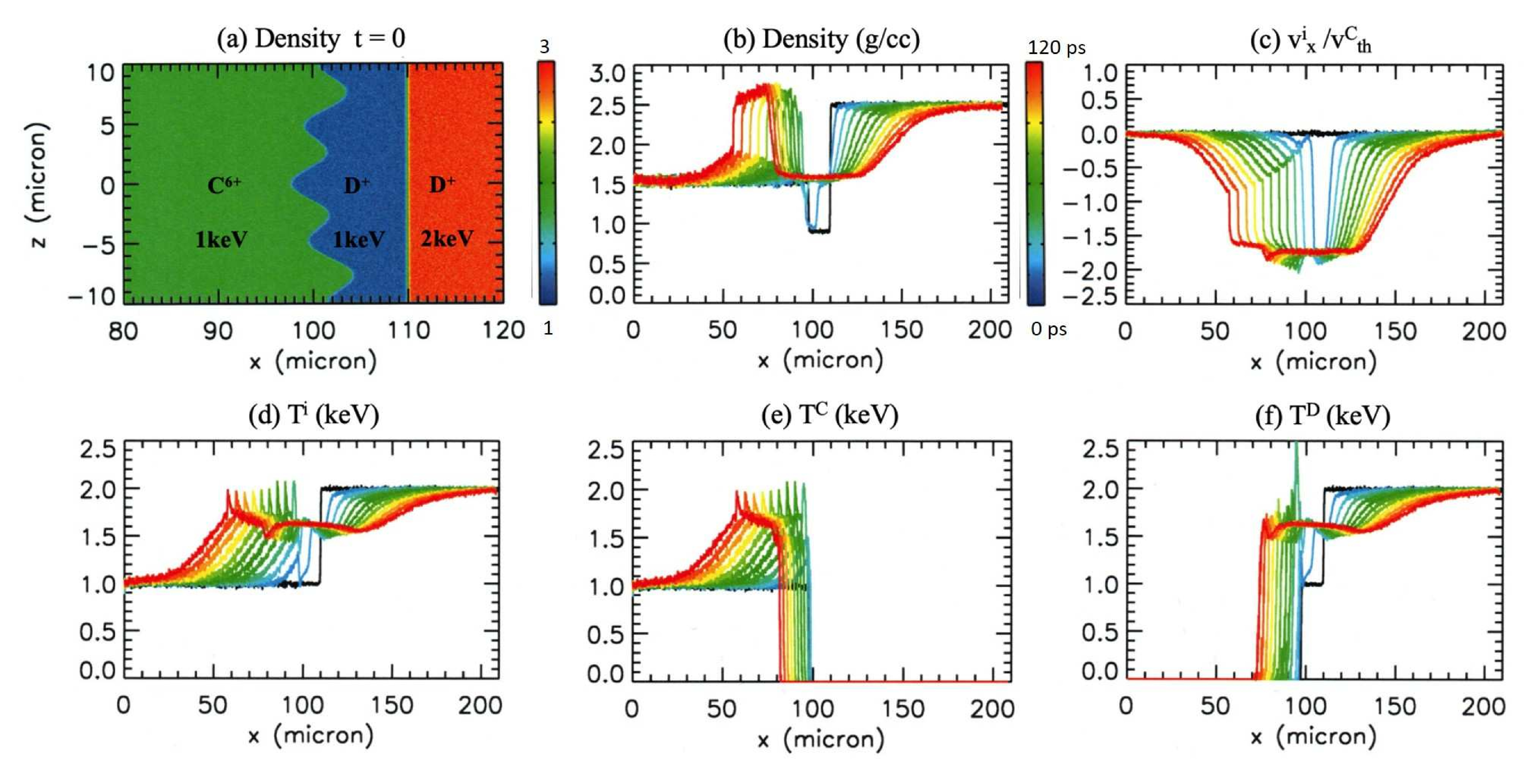}
\caption{\label{fig:5} 
Hybrid simulation of Richtmyer-Meshkov instability evolution resulting from the interaction of a C-D interface with an incident shock:
(a) Initial conditions and interface structure shown by the density, 
(b) to (f) profiles (at $z=0$) of 
density, 
mass-weighted ion flow velocity (in units of C ion thermal velocity $v^C_{\mathrm{th}}$),
mass-weighted ion temperature $T^i$,
$T^C$,
and $T^D$ from $t=0$ to 117~ps
(early- to late-time profiles are indicated by the spectrum of colors from back to red).
}
\end{figure*}

Finally, we examine the interface dynamics with the presence of a shock, as commonly encountered in ICF settings.
The initial conditions used in the hybrid simulation 
are from a \textsc{xRAGE} radiation-hydrodynamics simulation of an ICF implosion.
The simulation has sizes
$L_x = 210.0$~\microns and
$L_z = 20.0$~\micron.
We use Coulomb logarithms for self- and cross-species collisions based on Ref.~\cite{huba:1998}.
The C-D interface, as shown in Fig.~\ref{fig:5} (a), 
has 1~keV electron and ion temperatures
and sinusoidal perturbation wavelengths $10L_z, L_z, L_z/4$ with phase shifts
and with peak-to-peak amplitude $a=3.4$~\micron.
The C and D ions at the interface have densities 1.54 g/cc and 0.90 g/cc, respectively,
and the pressure balance across the interface is satisfied by
requiring $n_{\rm C} (1+Z_{\rm C}) = n_{\rm D} (1+Z_{\rm D})$.
The shock is initiated from a higher pressure region with D ions on the right side of the simulation box 
with density 2.5~g/cc and $T_e = T_D = 2$~keV.
The initial density profile is 
indicated by the black curve in frame (b).
With these plasma conditions, the total pressure jumps by 5.6 times across the boundary of D ions at 
different densities ($x=110$~\micron).

The time evolutions of the profiles (at $z=0$) of density, 
mass-weighted ion flow velocity,
mass-weighted ion temperature $T_i$,
$T_C$,
and $T_D$ 
from $t=0$ to 117~ps
are shown in frames (b) to (f) 
where the spectrum of colors from black to red is used to indicate early- to late-time profiles.
The shock drives a wide region of flowing plasma with speed $\sim 1.7$ times the C ion thermal speed. 
The shocked C material develops a high-density layer with a narrow width between $x \sim 55$ and 75~\microns
where the C ion temperature peaks at 2$\times$  the initial temperature; 
the temperature of the D ion layer at the interface is lower ($\sim 1.6$~keV).
The hybrid simulation provides  additional information on the energy partition between the two ion species.
The mass-weighted plasma density and temperature are in general agreement with those from the \textsc{xRAGE} simulation
with a plasma transport model. 

\section{Discussion}
\label{sec:summary}

We have presented the Hybrid-VPIC code, which extends the open-source VPIC code to include hybrid fluid/kinetic models. The code implements fairly standard hybrid solvers, and details of a few specific choices are documented for other code users. Sample applications from space and laboratory plasma modeling illustrated problems where the code has already proved useful, and they highlighted a couple of numerical complications that arise in hybrid PIC simulation. The new treatment of open boundaries, in particular, may be useful to other developers of hybrid PIC codes. 

Hybrid-VPIC also incorporates a relatively new electron fluid closure\cite{simakov:2014} for collisional unmagnetized regimes. Coupling this model to the hybrid PIC scheme has enabled studies of interfacial mix in HED settings, which is a very computationally demanding problem. A large computing effort has already begun on this application area, and some initial science results were described. In the future, the collision model may be updated with a new formulation \cite{higginson:2022}. In addition , the transport model for the electrons may be extended to include magnetic effects through an extension \cite{simakov:2022} of the Braginskii closure \cite{braginskii:1958}. This would allow additional effects including self-generated magnetic fields to be studied.

Finally, future work will also include updating the Hybrid-VPIC codebase to use routines from the newer VPIC 2.0 \cite{bird:2021} version, which is built on the Kokkos framework \cite{edwards:2014}. This framework allows high performance and portability on CPU-based, GPU-based, and other computing architectures.

\begin{acknowledgments}
Work performed under the auspices of the U.S.\ Department of Energy National Nuclear Security Administration under Contract No.\ 89233218CNA000001. Support for this work was provided, in part,
by the Advanced Simulation and Computing Integrated Computing Program as well as the Office of Experimental Sciences and Inertial Confinement Fusion Programs. Additional support was provided by NASA's Magnetospheric Multiscale (MMS) mission and the LANL Laboratory Directed Research and Development (LDRD) program. The authors thank Drs. Erik L. Vold and Jan Velechovsky
for useful discussions on the interface mix application and for providing the initial conditions from \textsc{xRAGE} radiation-hydrodynamics simulation
for the hybrid simulation in Fig. 8. 
VPIC simulations were run on ASC Trinity supercomputer under Capability
Class Computing and the Large Scale Calculations Initiative (LSCI) and on LANL Institutional Computing resources.
\end{acknowledgments}

\section*{Data Availability}
\label{s:data}
Data may be reproduced with the Hybrid-VPIC code, which is being released with open source code at  \url{https://github.com/lanl/vpic-kokkos/tree/hybridVPIC} \cite{hybridvpic}. Not all of the features described in the paper have been readied for release. Upon request to the authors, release will be expedited for features that other users need for particular studies.

\appendix
\section{A hybrid model with kinetic electrons} 
\label{sec:ebeam}
The main applications of the Hybrid-VPIC code have involved a kinetic ion model coupled to a fluid electron model. It is possible, however, to include a kinetic electron species in a hybrid PIC code. One example where this is crucial to study the stability of a beam of relativistic electrons propagating through air at atmospheric density. An MeV-range relativistic electron beam will partially ionize the background the air. The high-energy beam electrons may have mean-free paths of hundreds of meters. The background plasma, on the other hand, is highly collisional and remains relatively cool (a few eV). The kinetic length scales of the cold background plasma can be sub-micron, and it would not be feasible to handle this problem with a fully kinetic explicit PIC code. Rather, the cold background can be treated with a simplified fluid model.

A long wavelength instability of a relativistic electron beam propagating through a resistive background of air was treated first by \citet{rosenbluth:1960}. A simple model that captures the instability includes the collisionless Vlasov equation for the relativistic electrons and treats the air as a resistive background characterized by the simple Ohm's law ${\bf{E}} = \eta{\bf{J}}$ (where $\bf{J}$ is the current carried by the background). This model has been implemented in Hybrid-VPIC by treating the beam electrons as a kinetic species with a relativistic particle push and by reducing the Ohm's law to the simple form above, where ${\bf{J}} = \nabla\times{\bf{B}}/\mu_0 - {\bf{J}}_{eb}$ and ${\bf{J}}_{eb}$ is the current carried by the kinetic electron beam. The system is susceptible to an instability (now called the resistive hose instability), caused by the resistive diffusion of the magnetic field that allows the beam and magnetic field lines to slip past one another. The instability causes a nearly rigid transverse displacement of the the current channel, which corresponds to a kinking of the current for finite wavelengths.

\begin{figure}[h]
\includegraphics[width=166mm]{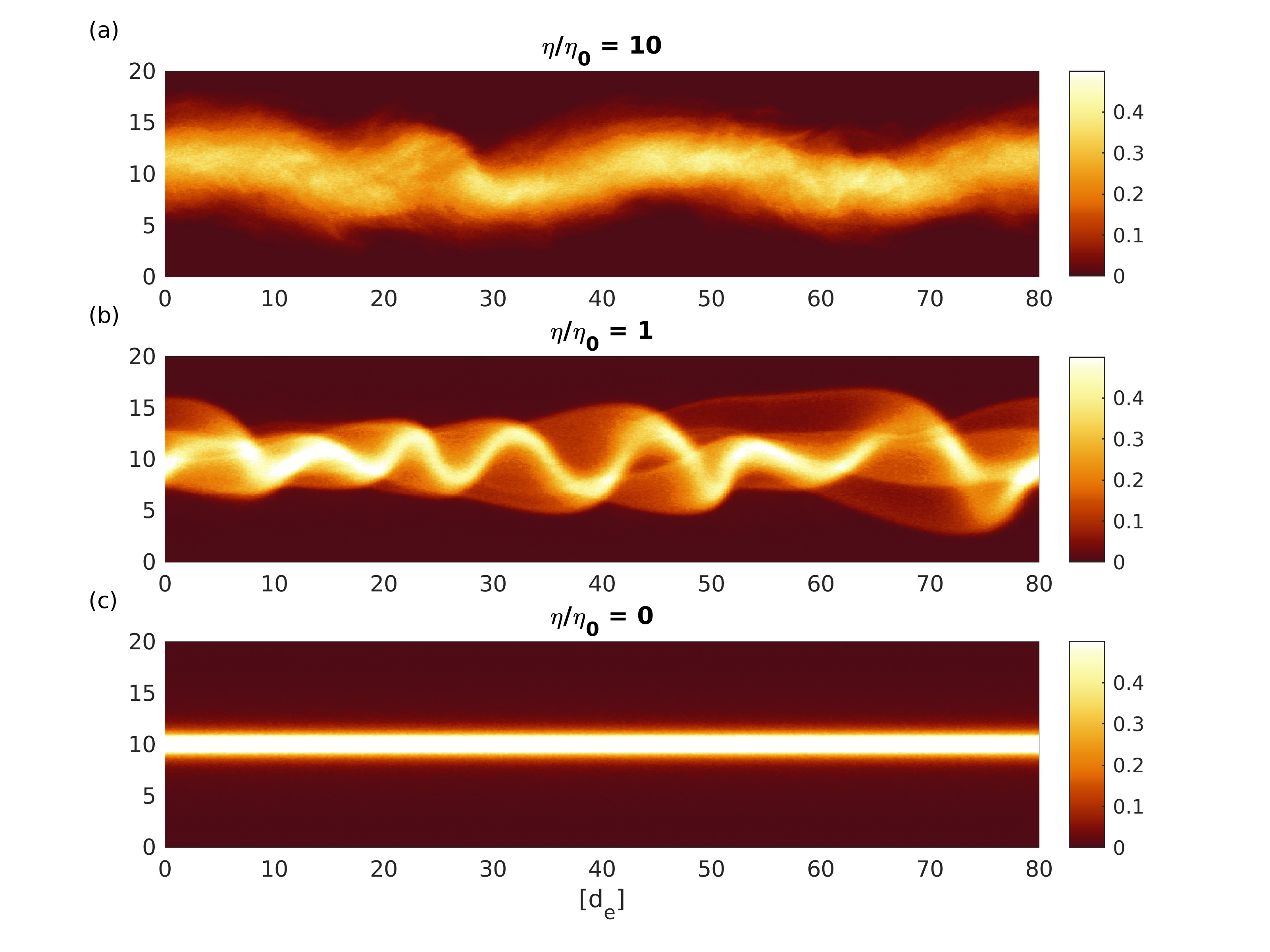}
\caption{\label{fig:ebeam} 
2D simulations of the resistive hose instability of a relativistic electron beam propagating through a resistive background. Each panel shows the electron density (normalized to the initial peak electron density $n_0$) at time $t=300/\omega_{pe}$, and distances are measured in electron skin depths $d_e=c/\omega_{pe}$ based on $n_0$. The resistivity is normalized to electron kinetic scales with a base unit of $\eta_0 = \mu_0c^2/\omega_{pe}$. (a) The most resistive background has the largest growth rate, and the electron beam is rapidly scattered and diffused. (b) The moderate resistivity case shows the transition of the linear instability into a strongly nonlinear regime. (c) A case without resistivity is shown for reference, and the electron beam retains an equilibrium profile.
}
\end{figure}

Example data are shown from simulations with with a $\sim1~MeV$ electron beam propagating through backgrounds with different resistivities. The beam is uniform in $x$ and given a Harris sheet profile, $n(z) \sim n_0\tanh(z/d_e)$ with $d_e$ the electron skin depth, and the magnetic field is consistent with the current carried by the electron beam. The electron beam particles have a uniform drift corresponding to an energy of $\sim1~MeV$ and a thermal spread consistent with the current channel width. This is not an exact equilibrium because the Harris sheet solution is non-relativistic, but the beam rapidly settles into a stable near-equilibrium (see Fig.~\ref{fig:ebeam}(c) without any resistivity or electric field). The simulations are not very computationally expensive, and they have a domain of $L_x\times L_z = 80~d_e\times20~d_e$ and is $640\times320$ cells with periodic boundaries in $x$. The $z$ boundaries are open for the fields and absorb particles. The simulations have 2000 particles per cell where the density peaks. Each simulation can be run in under a half hour on a single node of a multi-core (128-CPU) computing cluster. The field solver is sub-cycled to meet the CFL condition on the magnetic diffusion equation satisfied by $\bf{B}$ in this simple model.

As predicted by theory, the beam is unstable to a hose instability that kinks the current channel. The instability grows fastest for the highest resistivity plotted in Fig.~\ref{fig:ebeam}. While this simple model captures the essential features of the resistive hose instability, a more complete and realistic model would require treating the background air chemistry\cite{uhm:1980} and electron-air collisions. The example nevertheless shows how kinetic electrons may be incorporated into Hybrid-VPIC to solve another class of problems that cannot be treated with a fully kinetic code.

\section{Quadratic Sum (QS) Particle Shapes}
\label{sec:qsappend}
The QS scheme discussed in Sec.~\ref{sec:hybridem} accumulates the charge density of an ion macroparticle of weight $w_p$ belonging to species with charge $Z_pe$ in a cell with index $I\leftrightarrow (i,j,k)$ and relative cell coordinates $(dx,dy,dz)$ using:
\begin{equation}     
\begin{cases}
    n_{i,j,k} = (Z_pew_p/6V_c)\times(3 - dx^2 - dy^2 - dz^2), \\
    n_{i+1,j,k} = (Z_pew_p/12V_c)\times(1 + dx)(1+dx), \\
    n_{i,j+1,k} = (Z_pew_p/12V_c)\times(1 + dy)(1+dy), \\
    n_{i,j,k+1} = (Z_pew_p/12V_c)\times(1 + dz)(1+dz), \\
    n_{i-1,j,k} = (Z_pew_p/12V_c)\times(1 - dx)(1-dx), \\
    n_{i,j-1,k} = (Z_pew_p/12V_c)\times(1 - dy)(1-dy), \\
    n_{i,j,k-1} = (Z_pew_p/12V_c)\times(1 - dz)(1-dz), \\
\end{cases}
\label{eq:quad}
\end{equation}
with zero density for all other cells, and where $(dx,dy,dz)$ is evaluated at the half time step $t^{n+1/2}$. The ion currents are likewise accumulated with an additional factor of the particle velocity: ${\bf{J}}_{i,j,k} = Z_pen_{i,j,k}{\bf{u_p}}$. For the particle advance, the fields are interpolated to the particle position using pre-computed interpolation coefficients of each field component $F\in\{E_x,E_y,E_z,B_x,B_y,B_z\}$ of the form:
\begin{equation}     
\begin{cases}
    F^0_{i,j,k} = (1/12)\times(6F_{i,j,k} + F_{i+1,j,k} + F_{i,j+1,k} + F_{i,j,k+1} + F_{i-1,j,k} + F_{i,j-1,k} + F_{i,j,k-1}), \\
    F^x_{i,j,k} = (1/6)\times(F_{i+1,j,k} - F_{i-1,j,k}), \\
    F^y_{i,j,k} = (1/6)\times(F_{i,j+1,k} - F_{i,j-1,k}), \\
    F^z_{i,j,k} = (1/6)\times(F_{i,j,k+1} - F_{i,j,k-1}), \\
    F^{xx}_{i,j,k} = (1/12)\times(F_{i+1,j,k} + F_{i-1,j,k} - 2F_{i,j,k}), \\
    F^{yy}_{i,j,k} = (1/12)\times(F_{i,j+1,k} + F_{i,j-1,k} - 2F_{i,j,k}), \\
    F^{zz}_{i,j,k} = (1/12)\times(F_{i,j,k+1} + F_{i,j,k-1} - 2F_{i,j,k}). \\
\end{cases}
\label{eq:quadFI}
\end{equation}
In terms of the interpolation coefficients in Eq.~\ref{eq:quadFI}, the field used to push a particle at position ${\bf{x}}_p$ in cell $I\leftrightarrow (i,j,k)$ with relative cell coordinates $(dx,dy,dz)$ is:
\begin{equation}     
F(x_p) = F^0 + dx(F^x + dxF^{xx}) + dy(F^y + dyF^{yy}) + dz(F^z + dzF^{zz}), 
\label{eq:quadF}
\end{equation}
where each interpolant is evaluated in cell $I\leftrightarrow (i,j,k)$. Taken together, Eqs.~\ref{eq:quad} and \ref{eq:quadF} result in a consistent interpolation scheme free of particle self-forces.


\begin{thebibliography}{86}%
\makeatletter
\providecommand \@ifxundefined [1]{%
 \@ifx{#1\undefined}
}%
\providecommand \@ifnum [1]{%
 \ifnum #1\expandafter \@firstoftwo
 \else \expandafter \@secondoftwo
 \fi
}%
\providecommand \@ifx [1]{%
 \ifx #1\expandafter \@firstoftwo
 \else \expandafter \@secondoftwo
 \fi
}%
\providecommand \natexlab [1]{#1}%
\providecommand \enquote  [1]{``#1''}%
\providecommand \bibnamefont  [1]{#1}%
\providecommand \bibfnamefont [1]{#1}%
\providecommand \citenamefont [1]{#1}%
\providecommand \href@noop [0]{\@secondoftwo}%
\providecommand \href [0]{\begingroup \@sanitize@url \@href}%
\providecommand \@href[1]{\@@startlink{#1}\@@href}%
\providecommand \@@href[1]{\endgroup#1\@@endlink}%
\providecommand \@sanitize@url [0]{\catcode `\\12\catcode `\$12\catcode
  `\&12\catcode `\#12\catcode `\^12\catcode `\_12\catcode `\%12\relax}%
\providecommand \@@startlink[1]{}%
\providecommand \@@endlink[0]{}%
\providecommand \url  [0]{\begingroup\@sanitize@url \@url }%
\providecommand \@url [1]{\endgroup\@href {#1}{\urlprefix }}%
\providecommand \urlprefix  [0]{URL }%
\providecommand \Eprint [0]{\href }%
\providecommand \doibase [0]{http://dx.doi.org/}%
\providecommand \selectlanguage [0]{\@gobble}%
\providecommand \bibinfo  [0]{\@secondoftwo}%
\providecommand \bibfield  [0]{\@secondoftwo}%
\providecommand \translation [1]{[#1]}%
\providecommand \BibitemOpen [0]{}%
\providecommand \bibitemStop [0]{}%
\providecommand \bibitemNoStop [0]{.\EOS\space}%
\providecommand \EOS [0]{\spacefactor3000\relax}%
\providecommand \BibitemShut  [1]{\csname bibitem#1\endcsname}%
\let\auto@bib@innerbib\@empty
\bibitem [{\citenamefont {Bowers}\ \emph
  {et~al.}(2008{\natexlab{a}})\citenamefont {Bowers}, \citenamefont {Albright},
  \citenamefont {Yin}, \citenamefont {Bergen},\ and\ \citenamefont
  {Kwan}}]{bowers:2008}%
  \BibitemOpen
  \bibfield  {author} {\bibinfo {author} {\bibfnamefont {K.~J.}\ \bibnamefont
  {Bowers}}, \bibinfo {author} {\bibfnamefont {B.}~\bibnamefont {Albright}},
  \bibinfo {author} {\bibfnamefont {L.}~\bibnamefont {Yin}}, \bibinfo {author}
  {\bibfnamefont {B.}~\bibnamefont {Bergen}}, \ and\ \bibinfo {author}
  {\bibfnamefont {T.}~\bibnamefont {Kwan}},\ }\href@noop {} {\bibfield
  {journal} {\bibinfo  {journal} {Physics of Plasmas}\ }\textbf {\bibinfo
  {volume} {15}},\ \bibinfo {pages} {055703} (\bibinfo {year}
  {2008}{\natexlab{a}})}\BibitemShut {NoStop}%
\bibitem [{\citenamefont {Bowers}\ \emph
  {et~al.}(2008{\natexlab{b}})\citenamefont {Bowers}, \citenamefont {Albright},
  \citenamefont {Bergen}, \citenamefont {Yin}, \citenamefont {Barker},\ and\
  \citenamefont {Kerbyson}}]{bowers:2008b}%
  \BibitemOpen
  \bibfield  {author} {\bibinfo {author} {\bibfnamefont {K.~J.}\ \bibnamefont
  {Bowers}}, \bibinfo {author} {\bibfnamefont {B.~J.}\ \bibnamefont
  {Albright}}, \bibinfo {author} {\bibfnamefont {B.}~\bibnamefont {Bergen}},
  \bibinfo {author} {\bibfnamefont {L.}~\bibnamefont {Yin}}, \bibinfo {author}
  {\bibfnamefont {K.~J.}\ \bibnamefont {Barker}}, \ and\ \bibinfo {author}
  {\bibfnamefont {D.~J.}\ \bibnamefont {Kerbyson}},\ }in\ \href@noop {} {\emph
  {\bibinfo {booktitle} {SC'08: Proceedings of the 2008 ACM/IEEE conference on
  Supercomputing}}}\ (\bibinfo {organization} {IEEE},\ \bibinfo {year} {2008})\
  pp.\ \bibinfo {pages} {1--11}\BibitemShut {NoStop}%
\bibitem [{\citenamefont {Bowers}\ \emph {et~al.}(2009)\citenamefont {Bowers},
  \citenamefont {Albright}, \citenamefont {Yin}, \citenamefont {Daughton},
  \citenamefont {Roytershteyn}, \citenamefont {Bergen},\ and\ \citenamefont
  {Kwan}}]{bowers:2009}%
  \BibitemOpen
  \bibfield  {author} {\bibinfo {author} {\bibfnamefont {K.~J.}\ \bibnamefont
  {Bowers}}, \bibinfo {author} {\bibfnamefont {B.~J.}\ \bibnamefont
  {Albright}}, \bibinfo {author} {\bibfnamefont {L.}~\bibnamefont {Yin}},
  \bibinfo {author} {\bibfnamefont {W.}~\bibnamefont {Daughton}}, \bibinfo
  {author} {\bibfnamefont {V.}~\bibnamefont {Roytershteyn}}, \bibinfo {author}
  {\bibfnamefont {B.}~\bibnamefont {Bergen}}, \ and\ \bibinfo {author}
  {\bibfnamefont {T.}~\bibnamefont {Kwan}},\ }in\ \href@noop {} {\emph
  {\bibinfo {booktitle} {Journal of Physics: Conference Series}}},\ Vol.\
  \bibinfo {volume} {180}\ (\bibinfo {organization} {IOP Publishing},\ \bibinfo
  {year} {2009})\ p.\ \bibinfo {pages} {012055}\BibitemShut {NoStop}%
\bibitem [{\citenamefont {{Los Alamos National
  Laboratory}}(2022)}]{hybridvpic}%
  \BibitemOpen
  \bibfield  {author} {\bibinfo {author} {\bibnamefont {{Los Alamos National
  Laboratory}}},\ }\href@noop {} {\enquote {\bibinfo {title} {{Hybrid-VPIC},
  {https://github.com/lanl/vpic-kokkos/tree/hybridVPIC}},}\ } (\bibinfo {year}
  {2022})\BibitemShut {NoStop}%
\bibitem [{\citenamefont {Lipatov}(2002)}]{lipatov:2002}%
  \BibitemOpen
  \bibfield  {author} {\bibinfo {author} {\bibfnamefont {A.~S.}\ \bibnamefont
  {Lipatov}},\ }\href@noop {} {\emph {\bibinfo {title} {The hybrid multiscale
  simulation technology: an introduction with application to astrophysical and
  laboratory plasmas}}}\ (\bibinfo  {publisher} {Springer Science \& Business
  Media},\ \bibinfo {year} {2002})\BibitemShut {NoStop}%
\bibitem [{\citenamefont {Winske}\ \emph {et~al.}(2003)\citenamefont {Winske},
  \citenamefont {Yin}, \citenamefont {Omidi}, \citenamefont {Karimabadi},\ and\
  \citenamefont {Quest}}]{winske:2003}%
  \BibitemOpen
  \bibfield  {author} {\bibinfo {author} {\bibfnamefont {D.}~\bibnamefont
  {Winske}}, \bibinfo {author} {\bibfnamefont {L.}~\bibnamefont {Yin}},
  \bibinfo {author} {\bibfnamefont {N.}~\bibnamefont {Omidi}}, \bibinfo
  {author} {\bibfnamefont {H.}~\bibnamefont {Karimabadi}}, \ and\ \bibinfo
  {author} {\bibfnamefont {K.}~\bibnamefont {Quest}},\ }\href@noop {}
  {\bibfield  {journal} {\bibinfo  {journal} {Space plasma simulation}\ ,\
  \bibinfo {pages} {136}} (\bibinfo {year} {2003})}\BibitemShut {NoStop}%
\bibitem [{\citenamefont {Winske}\ \emph {et~al.}(2022)\citenamefont {Winske},
  \citenamefont {Karimabadi}, \citenamefont {Le}, \citenamefont {Omidi},
  \citenamefont {Roytershteyn},\ and\ \citenamefont {Stanier}}]{winske:2022}%
  \BibitemOpen
  \bibfield  {author} {\bibinfo {author} {\bibfnamefont {D.}~\bibnamefont
  {Winske}}, \bibinfo {author} {\bibfnamefont {H.}~\bibnamefont {Karimabadi}},
  \bibinfo {author} {\bibfnamefont {A.}~\bibnamefont {Le}}, \bibinfo {author}
  {\bibfnamefont {N.}~\bibnamefont {Omidi}}, \bibinfo {author} {\bibfnamefont
  {V.}~\bibnamefont {Roytershteyn}}, \ and\ \bibinfo {author} {\bibfnamefont
  {A.}~\bibnamefont {Stanier}},\ }\href@noop {} {\bibfield  {journal} {\bibinfo
   {journal} {arXiv preprint arXiv:2204.01676}\ } (\bibinfo {year}
  {2022})}\BibitemShut {NoStop}%
\bibitem [{\citenamefont {Nieter}\ and\ \citenamefont
  {Cary}(2004)}]{nieter:2004}%
  \BibitemOpen
  \bibfield  {author} {\bibinfo {author} {\bibfnamefont {C.}~\bibnamefont
  {Nieter}}\ and\ \bibinfo {author} {\bibfnamefont {J.~R.}\ \bibnamefont
  {Cary}},\ }\href@noop {} {\bibfield  {journal} {\bibinfo  {journal} {Journal
  of Computational Physics}\ }\textbf {\bibinfo {volume} {196}},\ \bibinfo
  {pages} {448} (\bibinfo {year} {2004})}\BibitemShut {NoStop}%
\bibitem [{\citenamefont {Gargat{\'e}}\ \emph {et~al.}(2007)\citenamefont
  {Gargat{\'e}}, \citenamefont {Bingham}, \citenamefont {Fonseca},\ and\
  \citenamefont {Silva}}]{gargate:2007}%
  \BibitemOpen
  \bibfield  {author} {\bibinfo {author} {\bibfnamefont {L.}~\bibnamefont
  {Gargat{\'e}}}, \bibinfo {author} {\bibfnamefont {R.}~\bibnamefont
  {Bingham}}, \bibinfo {author} {\bibfnamefont {R.~A.}\ \bibnamefont
  {Fonseca}}, \ and\ \bibinfo {author} {\bibfnamefont {L.~O.}\ \bibnamefont
  {Silva}},\ }\href@noop {} {\bibfield  {journal} {\bibinfo  {journal}
  {Computer physics communications}\ }\textbf {\bibinfo {volume} {176}},\
  \bibinfo {pages} {419} (\bibinfo {year} {2007})}\BibitemShut {NoStop}%
\bibitem [{\citenamefont {Karimabadi}\ \emph {et~al.}(2011)\citenamefont
  {Karimabadi}, \citenamefont {Loring}, \citenamefont {Vu}, \citenamefont
  {Omelchenko}, \citenamefont {Tatineni}, \citenamefont {Majumdar},
  \citenamefont {Ayachit},\ and\ \citenamefont {Geveci}}]{karimabadi:2011}%
  \BibitemOpen
  \bibfield  {author} {\bibinfo {author} {\bibfnamefont {H.}~\bibnamefont
  {Karimabadi}}, \bibinfo {author} {\bibfnamefont {B.}~\bibnamefont {Loring}},
  \bibinfo {author} {\bibfnamefont {H.}~\bibnamefont {Vu}}, \bibinfo {author}
  {\bibfnamefont {Y.}~\bibnamefont {Omelchenko}}, \bibinfo {author}
  {\bibfnamefont {M.}~\bibnamefont {Tatineni}}, \bibinfo {author}
  {\bibfnamefont {A.}~\bibnamefont {Majumdar}}, \bibinfo {author}
  {\bibfnamefont {U.}~\bibnamefont {Ayachit}}, \ and\ \bibinfo {author}
  {\bibfnamefont {B.}~\bibnamefont {Geveci}},\ }in\ \href@noop {} {\emph
  {\bibinfo {booktitle} {5th international conference of numerical modeling of
  space plasma flows (astronum 2010)}}},\ Vol.\ \bibinfo {volume} {444}\
  (\bibinfo {year} {2011})\ p.\ \bibinfo {pages} {281}\BibitemShut {NoStop}%
\bibitem [{\citenamefont {M{\"u}ller}\ \emph {et~al.}(2011)\citenamefont
  {M{\"u}ller}, \citenamefont {Simon}, \citenamefont {Motschmann},
  \citenamefont {Sch{\"u}le}, \citenamefont {Glassmeier},\ and\ \citenamefont
  {Pringle}}]{muller:2011}%
  \BibitemOpen
  \bibfield  {author} {\bibinfo {author} {\bibfnamefont {J.}~\bibnamefont
  {M{\"u}ller}}, \bibinfo {author} {\bibfnamefont {S.}~\bibnamefont {Simon}},
  \bibinfo {author} {\bibfnamefont {U.}~\bibnamefont {Motschmann}}, \bibinfo
  {author} {\bibfnamefont {J.}~\bibnamefont {Sch{\"u}le}}, \bibinfo {author}
  {\bibfnamefont {K.-H.}\ \bibnamefont {Glassmeier}}, \ and\ \bibinfo {author}
  {\bibfnamefont {G.~J.}\ \bibnamefont {Pringle}},\ }\href@noop {} {\bibfield
  {journal} {\bibinfo  {journal} {Computer Physics Communications}\ }\textbf
  {\bibinfo {volume} {182}},\ \bibinfo {pages} {946} (\bibinfo {year}
  {2011})}\BibitemShut {NoStop}%
\bibitem [{\citenamefont {Omelchenko}\ and\ \citenamefont
  {Karimabadi}(2012)}]{omelchenko:2012}%
  \BibitemOpen
  \bibfield  {author} {\bibinfo {author} {\bibfnamefont {Y.~A.}\ \bibnamefont
  {Omelchenko}}\ and\ \bibinfo {author} {\bibfnamefont {H.}~\bibnamefont
  {Karimabadi}},\ }\href@noop {} {\bibfield  {journal} {\bibinfo  {journal}
  {Journal of Computational Physics}\ }\textbf {\bibinfo {volume} {231}},\
  \bibinfo {pages} {1766} (\bibinfo {year} {2012})}\BibitemShut {NoStop}%
\bibitem [{\citenamefont {Kunz}, \citenamefont {Stone},\ and\ \citenamefont
  {Bai}(2014)}]{kunz:2014}%
  \BibitemOpen
  \bibfield  {author} {\bibinfo {author} {\bibfnamefont {M.~W.}\ \bibnamefont
  {Kunz}}, \bibinfo {author} {\bibfnamefont {J.~M.}\ \bibnamefont {Stone}}, \
  and\ \bibinfo {author} {\bibfnamefont {X.-N.}\ \bibnamefont {Bai}},\
  }\href@noop {} {\bibfield  {journal} {\bibinfo  {journal} {Journal of
  Computational Physics}\ }\textbf {\bibinfo {volume} {259}},\ \bibinfo {pages}
  {154} (\bibinfo {year} {2014})}\BibitemShut {NoStop}%
\bibitem [{\citenamefont {Fatemi}\ \emph {et~al.}(2017)\citenamefont {Fatemi},
  \citenamefont {Poppe}, \citenamefont {Delory},\ and\ \citenamefont
  {Farrell}}]{fatemi:2017}%
  \BibitemOpen
  \bibfield  {author} {\bibinfo {author} {\bibfnamefont {S.}~\bibnamefont
  {Fatemi}}, \bibinfo {author} {\bibfnamefont {A.~R.}\ \bibnamefont {Poppe}},
  \bibinfo {author} {\bibfnamefont {G.~T.}\ \bibnamefont {Delory}}, \ and\
  \bibinfo {author} {\bibfnamefont {W.~M.}\ \bibnamefont {Farrell}},\ }in\
  \href@noop {} {\emph {\bibinfo {booktitle} {Journal of Physics: Conference
  Series}}},\ Vol.\ \bibinfo {volume} {837}\ (\bibinfo {organization} {IOP
  Publishing},\ \bibinfo {year} {2017})\ p.\ \bibinfo {pages}
  {012017}\BibitemShut {NoStop}%
\bibitem [{\citenamefont {Peterson}, \citenamefont {Welch},\ and\ \citenamefont
  {Rose}(2018)}]{peterson:2018}%
  \BibitemOpen
  \bibfield  {author} {\bibinfo {author} {\bibfnamefont {K.}~\bibnamefont
  {Peterson}}, \bibinfo {author} {\bibfnamefont {D.}~\bibnamefont {Welch}}, \
  and\ \bibinfo {author} {\bibfnamefont {D.~V.}\ \bibnamefont {Rose}},\
  }\href@noop {} {\enquote {\bibinfo {title} {Pic and pic/fluid modeling in
  chicago: Algorithms and key computational issues for modeling on z today.}}\
  }\bibinfo {type} {Tech. Rep.}\ (\bibinfo  {institution} {Sandia National
  Lab.(SNL-NM), Albuquerque, NM (United States)},\ \bibinfo {year}
  {2018})\BibitemShut {NoStop}%
\bibitem [{\citenamefont {Haggerty}\ and\ \citenamefont
  {Caprioli}(2019)}]{haggerty:2019}%
  \BibitemOpen
  \bibfield  {author} {\bibinfo {author} {\bibfnamefont {C.~C.}\ \bibnamefont
  {Haggerty}}\ and\ \bibinfo {author} {\bibfnamefont {D.}~\bibnamefont
  {Caprioli}},\ }\href@noop {} {\bibfield  {journal} {\bibinfo  {journal} {The
  Astrophysical Journal}\ }\textbf {\bibinfo {volume} {887}},\ \bibinfo {pages}
  {165} (\bibinfo {year} {2019})}\BibitemShut {NoStop}%
\bibitem [{\citenamefont {Cohen}\ \emph {et~al.}(2019)\citenamefont {Cohen},
  \citenamefont {Larson}, \citenamefont {Belyaev},\ and\ \citenamefont
  {Thomas}}]{cohen:2019}%
  \BibitemOpen
  \bibfield  {author} {\bibinfo {author} {\bibfnamefont {B.}~\bibnamefont
  {Cohen}}, \bibinfo {author} {\bibfnamefont {D.}~\bibnamefont {Larson}},
  \bibinfo {author} {\bibfnamefont {M.}~\bibnamefont {Belyaev}}, \ and\
  \bibinfo {author} {\bibfnamefont {V.}~\bibnamefont {Thomas}},\ }\href@noop {}
  {\enquote {\bibinfo {title} {Topanga: A modern code for e3 simulations},}\
  }\bibinfo {type} {Tech. Rep.}\ (\bibinfo  {institution} {Lawrence Livermore
  National Lab.(LLNL), Livermore, CA (United States)},\ \bibinfo {year}
  {2019})\BibitemShut {NoStop}%
\bibitem [{\citenamefont {Le}\ \emph {et~al.}(2016{\natexlab{a}})\citenamefont
  {Le}, \citenamefont {Daughton}, \citenamefont {Karimabadi},\ and\
  \citenamefont {Egedal}}]{le:2016ani}%
  \BibitemOpen
  \bibfield  {author} {\bibinfo {author} {\bibfnamefont {A.}~\bibnamefont
  {Le}}, \bibinfo {author} {\bibfnamefont {W.}~\bibnamefont {Daughton}},
  \bibinfo {author} {\bibfnamefont {H.}~\bibnamefont {Karimabadi}}, \ and\
  \bibinfo {author} {\bibfnamefont {J.}~\bibnamefont {Egedal}},\ }\href@noop {}
  {\bibfield  {journal} {\bibinfo  {journal} {Physics of Plasmas}\ }\textbf
  {\bibinfo {volume} {23}},\ \bibinfo {pages} {032114} (\bibinfo {year}
  {2016}{\natexlab{a}})}\BibitemShut {NoStop}%
\bibitem [{\citenamefont {Pritchett}(2003)}]{pritchett:2003}%
  \BibitemOpen
  \bibfield  {author} {\bibinfo {author} {\bibfnamefont {P.~L.}\ \bibnamefont
  {Pritchett}},\ }\href@noop {} {\bibfield  {journal} {\bibinfo  {journal}
  {LECTURE NOTES IN PHYSICS-NEW YORK THEN BERLIN-}\ ,\ \bibinfo {pages} {1}}
  (\bibinfo {year} {2003})}\BibitemShut {NoStop}%
\bibitem [{\citenamefont {Stanier}, \citenamefont {Chac{\'o}n},\ and\
  \citenamefont {Chen}(2019)}]{stanier:2019}%
  \BibitemOpen
  \bibfield  {author} {\bibinfo {author} {\bibfnamefont {A.}~\bibnamefont
  {Stanier}}, \bibinfo {author} {\bibfnamefont {L.}~\bibnamefont {Chac{\'o}n}},
  \ and\ \bibinfo {author} {\bibfnamefont {G.}~\bibnamefont {Chen}},\
  }\href@noop {} {\bibfield  {journal} {\bibinfo  {journal} {Journal of
  Computational Physics}\ }\textbf {\bibinfo {volume} {376}},\ \bibinfo {pages}
  {597} (\bibinfo {year} {2019})}\BibitemShut {NoStop}%
\bibitem [{\citenamefont {Boris}\ \emph {et~al.}(1970)\citenamefont {Boris}
  \emph {et~al.}}]{boris:1970}%
  \BibitemOpen
  \bibfield  {author} {\bibinfo {author} {\bibfnamefont {J.~P.}\ \bibnamefont
  {Boris}},\ }in\ \href@noop {} {\emph {\bibinfo {booktitle}
  {Proc. Fourth Conf. Num. Sim. Plasmas}}}\ (\bibinfo {year} {1970})\ pp.\
  \bibinfo {pages} {3--67}\BibitemShut {NoStop}%
\bibitem [{\citenamefont {Harned}(1982)}]{harned:1982}%
  \BibitemOpen
  \bibfield  {author} {\bibinfo {author} {\bibfnamefont {D.~S.}\ \bibnamefont
  {Harned}},\ }\href@noop {} {\bibfield  {journal} {\bibinfo  {journal}
  {Journal of Computational Physics}\ }\textbf {\bibinfo {volume} {47}},\
  \bibinfo {pages} {452} (\bibinfo {year} {1982})}\BibitemShut {NoStop}%
\bibitem [{\citenamefont {Matthews}(1994)}]{matthews:1994}%
  \BibitemOpen
  \bibfield  {author} {\bibinfo {author} {\bibfnamefont {A.~P.}\ \bibnamefont
  {Matthews}},\ }\href@noop {} {\bibfield  {journal} {\bibinfo  {journal}
  {Journal of Computational Physics}\ }\textbf {\bibinfo {volume} {112}},\
  \bibinfo {pages} {102} (\bibinfo {year} {1994})}\BibitemShut {NoStop}%
\bibitem [{\citenamefont {Karimabadi}\ \emph {et~al.}(2004)\citenamefont
  {Karimabadi}, \citenamefont {Krauss-Varban}, \citenamefont {Huba},\ and\
  \citenamefont {Vu}}]{karimabadi:2004}%
  \BibitemOpen
  \bibfield  {author} {\bibinfo {author} {\bibfnamefont {H.}~\bibnamefont
  {Karimabadi}}, \bibinfo {author} {\bibfnamefont {D.}~\bibnamefont
  {Krauss-Varban}}, \bibinfo {author} {\bibfnamefont {J.}~\bibnamefont {Huba}},
  \ and\ \bibinfo {author} {\bibfnamefont {H.}~\bibnamefont {Vu}},\ }\href@noop
  {} {\bibfield  {journal} {\bibinfo  {journal} {Journal of Geophysical
  Research: Space Physics}\ }\textbf {\bibinfo {volume} {109}} (\bibinfo {year}
  {2004})}\BibitemShut {NoStop}%
\bibitem [{\citenamefont {Yee}(1966)}]{yee:1966}%
  \BibitemOpen
  \bibfield  {author} {\bibinfo {author} {\bibfnamefont {K.}~\bibnamefont
  {Yee}},\ }\href@noop {} {\bibfield  {journal} {\bibinfo  {journal} {IEEE
  Transactions on antennas and propagation}\ }\textbf {\bibinfo {volume}
  {14}},\ \bibinfo {pages} {302} (\bibinfo {year} {1966})}\BibitemShut
  {NoStop}%
\bibitem [{\citenamefont {Stanier}, \citenamefont {Chac\'{o}n},\ and\
  \citenamefont {Le}(2020)}]{stanier:2020}%
  \BibitemOpen
  \bibfield  {author} {\bibinfo {author} {\bibfnamefont {A.}~\bibnamefont
  {Stanier}}, \bibinfo {author} {\bibfnamefont {L.}~\bibnamefont {Chac\'{o}n}},
  \ and\ \bibinfo {author} {\bibfnamefont {A.}~\bibnamefont {Le}},\ }\href@noop
  {} {\bibfield  {journal} {\bibinfo  {journal} {Journal of Computational
  Physics}\ }\textbf {\bibinfo {volume} {420}},\ \bibinfo {pages} {109705}
  (\bibinfo {year} {2020})}\BibitemShut {NoStop}%
\bibitem [{\citenamefont {Daughton}, \citenamefont {Scudder},\ and\
  \citenamefont {Karimabadi}(2006)}]{daughton:2006}%
  \BibitemOpen
  \bibfield  {author} {\bibinfo {author} {\bibfnamefont {W.}~\bibnamefont
  {Daughton}}, \bibinfo {author} {\bibfnamefont {J.}~\bibnamefont {Scudder}}, \
  and\ \bibinfo {author} {\bibfnamefont {H.}~\bibnamefont {Karimabadi}},\
  }\href@noop {} {\bibfield  {journal} {\bibinfo  {journal} {Physics of
  Plasmas}\ }\textbf {\bibinfo {volume} {13}},\ \bibinfo {pages} {072101}
  (\bibinfo {year} {2006})}\BibitemShut {NoStop}%
\bibitem [{\citenamefont {Swift}(1995)}]{swift:1995}%
  \BibitemOpen
  \bibfield  {author} {\bibinfo {author} {\bibfnamefont {D.~W.}\ \bibnamefont
  {Swift}},\ }\href@noop {} {\bibfield  {journal} {\bibinfo  {journal}
  {Geophysical research letters}\ }\textbf {\bibinfo {volume} {22}},\ \bibinfo
  {pages} {311} (\bibinfo {year} {1995})}\BibitemShut {NoStop}%
\bibitem [{\citenamefont {Karimabadi}\ \emph {et~al.}(2006)\citenamefont
  {Karimabadi}, \citenamefont {Vu}, \citenamefont {Krauss-Varban},\ and\
  \citenamefont {Omelchenko}}]{karimabadi:2006}%
  \BibitemOpen
  \bibfield  {author} {\bibinfo {author} {\bibfnamefont {H.}~\bibnamefont
  {Karimabadi}}, \bibinfo {author} {\bibfnamefont {H.}~\bibnamefont {Vu}},
  \bibinfo {author} {\bibfnamefont {D.}~\bibnamefont {Krauss-Varban}}, \ and\
  \bibinfo {author} {\bibfnamefont {Y.}~\bibnamefont {Omelchenko}},\ }in\
  \href@noop {} {\emph {\bibinfo {booktitle} {Numerical modeling of Space
  plasma flows}}},\ Vol.\ \bibinfo {volume} {359}\ (\bibinfo {year} {2006})\
  p.\ \bibinfo {pages} {257}\BibitemShut {NoStop}%
\bibitem [{\citenamefont {Tr{\'a}vn{\'\i}{\v{c}}ek}, \citenamefont
  {Hellinger},\ and\ \citenamefont {Schriver}(2007)}]{travnivcek:2007}%
  \BibitemOpen
  \bibfield  {author} {\bibinfo {author} {\bibfnamefont {P.}~\bibnamefont
  {Tr{\'a}vn{\'\i}{\v{c}}ek}}, \bibinfo {author} {\bibfnamefont
  {P.}~\bibnamefont {Hellinger}}, \ and\ \bibinfo {author} {\bibfnamefont
  {D.}~\bibnamefont {Schriver}},\ }\href@noop {} {\bibfield  {journal}
  {\bibinfo  {journal} {Geophysical research letters}\ }\textbf {\bibinfo
  {volume} {34}} (\bibinfo {year} {2007})}\BibitemShut {NoStop}%
\bibitem [{\citenamefont {Omidi}, \citenamefont {Eastwood},\ and\ \citenamefont
  {Sibeck}(2010)}]{omidi:2010}%
  \BibitemOpen
  \bibfield  {author} {\bibinfo {author} {\bibfnamefont {N.}~\bibnamefont
  {Omidi}}, \bibinfo {author} {\bibfnamefont {J.}~\bibnamefont {Eastwood}}, \
  and\ \bibinfo {author} {\bibfnamefont {D.}~\bibnamefont {Sibeck}},\
  }\href@noop {} {\bibfield  {journal} {\bibinfo  {journal} {Journal of
  Geophysical Research: Space Physics}\ }\textbf {\bibinfo {volume} {115}}
  (\bibinfo {year} {2010})}\BibitemShut {NoStop}%
\bibitem [{\citenamefont {Lin}\ \emph {et~al.}(2014)\citenamefont {Lin},
  \citenamefont {Wang}, \citenamefont {Lu}, \citenamefont {Perez},\ and\
  \citenamefont {Lu}}]{lin:2014}%
  \BibitemOpen
  \bibfield  {author} {\bibinfo {author} {\bibfnamefont {Y.}~\bibnamefont
  {Lin}}, \bibinfo {author} {\bibfnamefont {X.}~\bibnamefont {Wang}}, \bibinfo
  {author} {\bibfnamefont {S.}~\bibnamefont {Lu}}, \bibinfo {author}
  {\bibfnamefont {J.}~\bibnamefont {Perez}}, \ and\ \bibinfo {author}
  {\bibfnamefont {Q.}~\bibnamefont {Lu}},\ }\href@noop {} {\bibfield  {journal}
  {\bibinfo  {journal} {Journal of Geophysical Research: Space Physics}\
  }\textbf {\bibinfo {volume} {119}},\ \bibinfo {pages} {7413} (\bibinfo {year}
  {2014})}\BibitemShut {NoStop}%
\bibitem [{\citenamefont {Klein}\ and\ \citenamefont
  {Vech}(2019)}]{Klein:2019}%
  \BibitemOpen
  \bibfield  {author} {\bibinfo {author} {\bibfnamefont {K.~G.}\ \bibnamefont
  {Klein}}\ and\ \bibinfo {author} {\bibfnamefont {D.}~\bibnamefont {Vech}},\
  }\href {\doibase 10.3847/2515-5172/ab3465} {\bibfield  {journal} {\bibinfo
  {journal} {Research Notes of the AAS}\ }\textbf {\bibinfo {volume} {3}},\
  \bibinfo {pages} {107} (\bibinfo {year} {2019})}\BibitemShut {NoStop}%
\bibitem [{\citenamefont {Gary}(1991)}]{gary:1991}%
  \BibitemOpen
  \bibfield  {author} {\bibinfo {author} {\bibfnamefont {S.~P.}\ \bibnamefont
  {Gary}},\ }\href@noop {} {\bibfield  {journal} {\bibinfo  {journal} {Space
  Science Reviews}\ }\textbf {\bibinfo {volume} {56}},\ \bibinfo {pages} {373}
  (\bibinfo {year} {1991})}\BibitemShut {NoStop}%
\bibitem [{\citenamefont {Keenan}\ \emph {et~al.}(2022)\citenamefont {Keenan},
  \citenamefont {Le}, \citenamefont {Winske}, \citenamefont {Stanier},
  \citenamefont {Wetherton}, \citenamefont {Cowee},\ and\ \citenamefont
  {Guo}}]{keenan:2022}%
  \BibitemOpen
  \bibfield  {author} {\bibinfo {author} {\bibfnamefont {B.~D.}\ \bibnamefont
  {Keenan}}, \bibinfo {author} {\bibfnamefont {A.}~\bibnamefont {Le}}, \bibinfo
  {author} {\bibfnamefont {D.}~\bibnamefont {Winske}}, \bibinfo {author}
  {\bibfnamefont {A.}~\bibnamefont {Stanier}}, \bibinfo {author} {\bibfnamefont
  {B.}~\bibnamefont {Wetherton}}, \bibinfo {author} {\bibfnamefont
  {M.}~\bibnamefont {Cowee}}, \ and\ \bibinfo {author} {\bibfnamefont
  {F.}~\bibnamefont {Guo}},\ }\href@noop {} {\bibfield  {journal} {\bibinfo
  {journal} {Physics of Plasmas}\ }\textbf {\bibinfo {volume} {29}},\ \bibinfo
  {pages} {012107} (\bibinfo {year} {2022})}\BibitemShut {NoStop}%
\bibitem [{\citenamefont {Le}\ \emph {et~al.}(2023)\citenamefont {Le},
  \citenamefont {Chen}, \citenamefont {Wetherton}, \citenamefont {Keenan},\
  and\ \citenamefont {Stanier}}]{le:2023}%
  \BibitemOpen
  \bibfield  {author} {\bibinfo {author} {\bibfnamefont {A.}~\bibnamefont
  {Le}}, \bibinfo {author} {\bibfnamefont {L.-J.}\ \bibnamefont {Chen}},
  \bibinfo {author} {\bibfnamefont {B.}~\bibnamefont {Wetherton}}, \bibinfo
  {author} {\bibfnamefont {B.}~\bibnamefont {Keenan}}, \ and\ \bibinfo {author}
  {\bibfnamefont {A.}~\bibnamefont {Stanier}},\ }\href {\doibase
  10.3389/fspas.2022.1100472} {\bibfield  {journal} {\bibinfo  {journal}
  {Frontiers in Astronomy and Space Sciences}\ }\textbf {\bibinfo {volume} {9}}
  (\bibinfo {year} {2023}),\ 10.3389/fspas.2022.1100472}\BibitemShut {NoStop}%
\bibitem [{\citenamefont {Winske}\ and\ \citenamefont
  {Gary}(2007)}]{winske:2007}%
  \BibitemOpen
  \bibfield  {author} {\bibinfo {author} {\bibfnamefont {D.}~\bibnamefont
  {Winske}}\ and\ \bibinfo {author} {\bibfnamefont {S.~P.}\ \bibnamefont
  {Gary}},\ }\href@noop {} {\bibfield  {journal} {\bibinfo  {journal} {Journal
  of Geophysical Research: Space Physics}\ }\textbf {\bibinfo {volume} {112}}
  (\bibinfo {year} {2007})}\BibitemShut {NoStop}%
\bibitem [{\citenamefont {Clark}\ \emph {et~al.}(2013)\citenamefont {Clark},
  \citenamefont {Winske}, \citenamefont {Schaeffer}, \citenamefont {Everson},
  \citenamefont {Bondarenko}, \citenamefont {Constantin},\ and\ \citenamefont
  {Niemann}}]{clark:2013}%
  \BibitemOpen
  \bibfield  {author} {\bibinfo {author} {\bibfnamefont {S.}~\bibnamefont
  {Clark}}, \bibinfo {author} {\bibfnamefont {D.}~\bibnamefont {Winske}},
  \bibinfo {author} {\bibfnamefont {D.}~\bibnamefont {Schaeffer}}, \bibinfo
  {author} {\bibfnamefont {E.}~\bibnamefont {Everson}}, \bibinfo {author}
  {\bibfnamefont {A.}~\bibnamefont {Bondarenko}}, \bibinfo {author}
  {\bibfnamefont {C.}~\bibnamefont {Constantin}}, \ and\ \bibinfo {author}
  {\bibfnamefont {C.}~\bibnamefont {Niemann}},\ }\href@noop {} {\bibfield
  {journal} {\bibinfo  {journal} {Physics of Plasmas}\ }\textbf {\bibinfo
  {volume} {20}},\ \bibinfo {pages} {082129} (\bibinfo {year}
  {2013})}\BibitemShut {NoStop}%
\bibitem [{\citenamefont {Winske}\ \emph {et~al.}(2019)\citenamefont {Winske},
  \citenamefont {Huba}, \citenamefont {Niemann},\ and\ \citenamefont
  {Le}}]{winske:2019}%
  \BibitemOpen
  \bibfield  {author} {\bibinfo {author} {\bibfnamefont {D.}~\bibnamefont
  {Winske}}, \bibinfo {author} {\bibfnamefont {J.~D.}\ \bibnamefont {Huba}},
  \bibinfo {author} {\bibfnamefont {C.}~\bibnamefont {Niemann}}, \ and\
  \bibinfo {author} {\bibfnamefont {A.}~\bibnamefont {Le}},\ }\href@noop {}
  {\bibfield  {journal} {\bibinfo  {journal} {Frontiers in Astronomy and Space
  Sciences}\ }\textbf {\bibinfo {volume} {5}},\ \bibinfo {pages} {51} (\bibinfo
  {year} {2019})}\BibitemShut {NoStop}%
\bibitem [{\citenamefont {Hewett}, \citenamefont {Brecht},\ and\ \citenamefont
  {Larson}(2011)}]{hewett:2011}%
  \BibitemOpen
  \bibfield  {author} {\bibinfo {author} {\bibfnamefont {D.~W.}\ \bibnamefont
  {Hewett}}, \bibinfo {author} {\bibfnamefont {S.~H.}\ \bibnamefont {Brecht}},
  \ and\ \bibinfo {author} {\bibfnamefont {D.~J.}\ \bibnamefont {Larson}},\
  }\href@noop {} {\bibfield  {journal} {\bibinfo  {journal} {Journal of
  Geophysical Research: Space Physics}\ }\textbf {\bibinfo {volume} {116}}
  (\bibinfo {year} {2011})}\BibitemShut {NoStop}%
\bibitem [{\citenamefont {Le}\ \emph {et~al.}(2021)\citenamefont {Le},
  \citenamefont {Winske}, \citenamefont {Stanier}, \citenamefont {Daughton},
  \citenamefont {Cowee}, \citenamefont {Wetherton},\ and\ \citenamefont
  {Guo}}]{le:2021}%
  \BibitemOpen
  \bibfield  {author} {\bibinfo {author} {\bibfnamefont {A.}~\bibnamefont
  {Le}}, \bibinfo {author} {\bibfnamefont {D.}~\bibnamefont {Winske}}, \bibinfo
  {author} {\bibfnamefont {A.}~\bibnamefont {Stanier}}, \bibinfo {author}
  {\bibfnamefont {W.}~\bibnamefont {Daughton}}, \bibinfo {author}
  {\bibfnamefont {M.}~\bibnamefont {Cowee}}, \bibinfo {author} {\bibfnamefont
  {B.}~\bibnamefont {Wetherton}}, \ and\ \bibinfo {author} {\bibfnamefont
  {F.}~\bibnamefont {Guo}},\ }\href@noop {} {\bibfield  {journal} {\bibinfo
  {journal} {Journal of Geophysical Research: Space Physics}\ }\textbf
  {\bibinfo {volume} {126}},\ \bibinfo {pages} {e2021JA029125} (\bibinfo {year}
  {2021})}\BibitemShut {NoStop}%
\bibitem [{\citenamefont {Ivanov}\ \emph {et~al.}(2003)\citenamefont {Ivanov},
  \citenamefont {Anikeev}, \citenamefont {Bagryansky}, \citenamefont
  {Deichuli}, \citenamefont {Korepanov}, \citenamefont {Lizunov}, \citenamefont
  {Maximov}, \citenamefont {Murakhtin}, \citenamefont {Savkin}, \citenamefont
  {Den~Hartog} \emph {et~al.}}]{ivanov:2003}%
  \BibitemOpen
  \bibfield  {author} {\bibinfo {author} {\bibfnamefont {A.}~\bibnamefont
  {Ivanov}}, \bibinfo {author} {\bibfnamefont {A.}~\bibnamefont {Anikeev}},
  \bibinfo {author} {\bibfnamefont {P.}~\bibnamefont {Bagryansky}}, \bibinfo
  {author} {\bibfnamefont {P.}~\bibnamefont {Deichuli}}, \bibinfo {author}
  {\bibfnamefont {S.}~\bibnamefont {Korepanov}}, \bibinfo {author}
  {\bibfnamefont {A.}~\bibnamefont {Lizunov}}, \bibinfo {author} {\bibfnamefont
  {V.}~\bibnamefont {Maximov}}, \bibinfo {author} {\bibfnamefont
  {S.}~\bibnamefont {Murakhtin}}, \bibinfo {author} {\bibfnamefont {V.~Y.}\
  \bibnamefont {Savkin}}, \bibinfo {author} {\bibfnamefont {D.}~\bibnamefont
  {Den~Hartog}},  \emph {et~al.},\ }\href@noop {} {\bibfield  {journal}
  {\bibinfo  {journal} {Physical review letters}\ }\textbf {\bibinfo {volume}
  {90}},\ \bibinfo {pages} {105002} (\bibinfo {year} {2003})}\BibitemShut
  {NoStop}%
\bibitem [{\citenamefont {Ivanov}\ and\ \citenamefont
  {Prikhodko}(2013)}]{ivanov:2013}%
  \BibitemOpen
  \bibfield  {author} {\bibinfo {author} {\bibfnamefont {A.~A.}\ \bibnamefont
  {Ivanov}}\ and\ \bibinfo {author} {\bibfnamefont {V.}~\bibnamefont
  {Prikhodko}},\ }\href@noop {} {\bibfield  {journal} {\bibinfo  {journal}
  {Plasma Physics and Controlled Fusion}\ }\textbf {\bibinfo {volume} {55}},\
  \bibinfo {pages} {063001} (\bibinfo {year} {2013})}\BibitemShut {NoStop}%
\bibitem [{\citenamefont {Wetherton}\ \emph {et~al.}(2021)\citenamefont
  {Wetherton}, \citenamefont {Le}, \citenamefont {Egedal}, \citenamefont
  {Forest}, \citenamefont {Daughton}, \citenamefont {Stanier},\ and\
  \citenamefont {Boldyrev}}]{wetherton:2021}%
  \BibitemOpen
  \bibfield  {author} {\bibinfo {author} {\bibfnamefont {B.~A.}\ \bibnamefont
  {Wetherton}}, \bibinfo {author} {\bibfnamefont {A.}~\bibnamefont {Le}},
  \bibinfo {author} {\bibfnamefont {J.}~\bibnamefont {Egedal}}, \bibinfo
  {author} {\bibfnamefont {C.}~\bibnamefont {Forest}}, \bibinfo {author}
  {\bibfnamefont {W.}~\bibnamefont {Daughton}}, \bibinfo {author}
  {\bibfnamefont {A.}~\bibnamefont {Stanier}}, \ and\ \bibinfo {author}
  {\bibfnamefont {S.}~\bibnamefont {Boldyrev}},\ }\href@noop {} {\bibfield
  {journal} {\bibinfo  {journal} {Physics of Plasmas}\ }\textbf {\bibinfo
  {volume} {28}},\ \bibinfo {pages} {042510} (\bibinfo {year}
  {2021})}\BibitemShut {NoStop}%
\bibitem [{\citenamefont {Bellei}\ \emph {et~al.}(2013)\citenamefont {Bellei},
  \citenamefont {Amendt}, \citenamefont {Wilks}, \citenamefont {Haines},
  \citenamefont {Casey}, \citenamefont {Li}, \citenamefont {Petrasso},\ and\
  \citenamefont {Welch}}]{bellei:2013}%
  \BibitemOpen
  \bibfield  {author} {\bibinfo {author} {\bibfnamefont {C.}~\bibnamefont
  {Bellei}}, \bibinfo {author} {\bibfnamefont {P.}~\bibnamefont {Amendt}},
  \bibinfo {author} {\bibfnamefont {S.}~\bibnamefont {Wilks}}, \bibinfo
  {author} {\bibfnamefont {M.}~\bibnamefont {Haines}}, \bibinfo {author}
  {\bibfnamefont {D.}~\bibnamefont {Casey}}, \bibinfo {author} {\bibfnamefont
  {C.}~\bibnamefont {Li}}, \bibinfo {author} {\bibfnamefont {R.}~\bibnamefont
  {Petrasso}}, \ and\ \bibinfo {author} {\bibfnamefont {D.}~\bibnamefont
  {Welch}},\ }\href@noop {} {\bibfield  {journal} {\bibinfo  {journal} {Physics
  of Plasmas}\ }\textbf {\bibinfo {volume} {20}},\ \bibinfo {pages} {012701}
  (\bibinfo {year} {2013})}\BibitemShut {NoStop}%
\bibitem [{\citenamefont {Le}\ \emph {et~al.}(2016{\natexlab{b}})\citenamefont
  {Le}, \citenamefont {Kwan}, \citenamefont {Schmitt}, \citenamefont
  {Herrmann},\ and\ \citenamefont {Batha}}]{le:2016}%
  \BibitemOpen
  \bibfield  {author} {\bibinfo {author} {\bibfnamefont {A.}~\bibnamefont
  {Le}}, \bibinfo {author} {\bibfnamefont {T.~J.}\ \bibnamefont {Kwan}},
  \bibinfo {author} {\bibfnamefont {M.~J.}\ \bibnamefont {Schmitt}}, \bibinfo
  {author} {\bibfnamefont {H.~W.}\ \bibnamefont {Herrmann}}, \ and\ \bibinfo
  {author} {\bibfnamefont {S.~H.}\ \bibnamefont {Batha}},\ }\href@noop {}
  {\bibfield  {journal} {\bibinfo  {journal} {Physics of Plasmas}\ }\textbf
  {\bibinfo {volume} {23}},\ \bibinfo {pages} {102705} (\bibinfo {year}
  {2016}{\natexlab{b}})}\BibitemShut {NoStop}%
\bibitem [{\citenamefont {Sio}\ \emph {et~al.}(2019)\citenamefont {Sio},
  \citenamefont {Frenje}, \citenamefont {Le}, \citenamefont {Atzeni},
  \citenamefont {Kwan}, \citenamefont {Johnson}, \citenamefont {Kagan},
  \citenamefont {Stoeckl}, \citenamefont {Li}, \citenamefont {Parker} \emph
  {et~al.}}]{sio:2019}%
  \BibitemOpen
  \bibfield  {author} {\bibinfo {author} {\bibfnamefont {H.}~\bibnamefont
  {Sio}}, \bibinfo {author} {\bibfnamefont {J.}~\bibnamefont {Frenje}},
  \bibinfo {author} {\bibfnamefont {A.}~\bibnamefont {Le}}, \bibinfo {author}
  {\bibfnamefont {S.}~\bibnamefont {Atzeni}}, \bibinfo {author} {\bibfnamefont
  {T.~J.}\ \bibnamefont {Kwan}}, \bibinfo {author} {\bibfnamefont {M.~G.}\
  \bibnamefont {Johnson}}, \bibinfo {author} {\bibfnamefont {G.}~\bibnamefont
  {Kagan}}, \bibinfo {author} {\bibfnamefont {C.}~\bibnamefont {Stoeckl}},
  \bibinfo {author} {\bibfnamefont {C.}~\bibnamefont {Li}}, \bibinfo {author}
  {\bibfnamefont {C.}~\bibnamefont {Parker}},  \emph {et~al.},\ }\href@noop {}
  {\bibfield  {journal} {\bibinfo  {journal} {Physical Review Letters}\
  }\textbf {\bibinfo {volume} {122}},\ \bibinfo {pages} {035001} (\bibinfo
  {year} {2019})}\BibitemShut {NoStop}%
\bibitem [{\citenamefont {Simakov}\ and\ \citenamefont
  {Molvig}(2014)}]{simakov:2014}%
  \BibitemOpen
  \bibfield  {author} {\bibinfo {author} {\bibfnamefont {A.~N.}\ \bibnamefont
  {Simakov}}\ and\ \bibinfo {author} {\bibfnamefont {K.}~\bibnamefont
  {Molvig}},\ }\href@noop {} {\bibfield  {journal} {\bibinfo  {journal}
  {Physics of Plasmas}\ }\textbf {\bibinfo {volume} {21}},\ \bibinfo {pages}
  {024503} (\bibinfo {year} {2014})}\BibitemShut {NoStop}%
\bibitem [{\citenamefont {Walsh}\ \emph {et~al.}(2017)\citenamefont {Walsh},
  \citenamefont {Chittenden}, \citenamefont {McGlinchey}, \citenamefont
  {Niasse},\ and\ \citenamefont {Appelbe}}]{walsh:2017}%
  \BibitemOpen
  \bibfield  {author} {\bibinfo {author} {\bibfnamefont {C.}~\bibnamefont
  {Walsh}}, \bibinfo {author} {\bibfnamefont {J.}~\bibnamefont {Chittenden}},
  \bibinfo {author} {\bibfnamefont {K.}~\bibnamefont {McGlinchey}}, \bibinfo
  {author} {\bibfnamefont {N.}~\bibnamefont {Niasse}}, \ and\ \bibinfo {author}
  {\bibfnamefont {B.}~\bibnamefont {Appelbe}},\ }\href@noop {} {\bibfield
  {journal} {\bibinfo  {journal} {Physical review letters}\ }\textbf {\bibinfo
  {volume} {118}},\ \bibinfo {pages} {155001} (\bibinfo {year}
  {2017})}\BibitemShut {NoStop}%
\bibitem [{\citenamefont {Sadler}, \citenamefont {Li},\ and\ \citenamefont
  {Flippo}(2020)}]{sadler:2020}%
  \BibitemOpen
  \bibfield  {author} {\bibinfo {author} {\bibfnamefont {J.~D.}\ \bibnamefont
  {Sadler}}, \bibinfo {author} {\bibfnamefont {H.}~\bibnamefont {Li}}, \ and\
  \bibinfo {author} {\bibfnamefont {K.~A.}\ \bibnamefont {Flippo}},\
  }\href@noop {} {\bibfield  {journal} {\bibinfo  {journal} {Philosophical
  Transactions of the Royal Society A}\ }\textbf {\bibinfo {volume} {378}},\
  \bibinfo {pages} {20200045} (\bibinfo {year} {2020})}\BibitemShut {NoStop}%
\bibitem [{\citenamefont {Takizuka}\ and\ \citenamefont
  {Abe}(1977)}]{takizuka:1977}%
  \BibitemOpen
  \bibfield  {author} {\bibinfo {author} {\bibfnamefont {T.}~\bibnamefont
  {Takizuka}}\ and\ \bibinfo {author} {\bibfnamefont {H.}~\bibnamefont {Abe}},\
  }\href@noop {} {\bibfield  {journal} {\bibinfo  {journal} {Journal of
  computational physics}\ }\textbf {\bibinfo {volume} {25}},\ \bibinfo {pages}
  {205} (\bibinfo {year} {1977})}\BibitemShut {NoStop}%
\bibitem [{\citenamefont {Daughton}\ \emph {et~al.}(2009)\citenamefont
  {Daughton}, \citenamefont {Roytershteyn}, \citenamefont {Albright},
  \citenamefont {Karimabadi}, \citenamefont {Yin},\ and\ \citenamefont
  {Bowers}}]{daughton:2009}%
  \BibitemOpen
  \bibfield  {author} {\bibinfo {author} {\bibfnamefont {W.}~\bibnamefont
  {Daughton}}, \bibinfo {author} {\bibfnamefont {V.}~\bibnamefont
  {Roytershteyn}}, \bibinfo {author} {\bibfnamefont {B.}~\bibnamefont
  {Albright}}, \bibinfo {author} {\bibfnamefont {H.}~\bibnamefont
  {Karimabadi}}, \bibinfo {author} {\bibfnamefont {L.}~\bibnamefont {Yin}}, \
  and\ \bibinfo {author} {\bibfnamefont {K.~J.}\ \bibnamefont {Bowers}},\
  }\href@noop {} {\bibfield  {journal} {\bibinfo  {journal} {Physical review
  letters}\ }\textbf {\bibinfo {volume} {103}},\ \bibinfo {pages} {065004}
  (\bibinfo {year} {2009})}\BibitemShut {NoStop}%
\bibitem [{\citenamefont {Roytershteyn}\ \emph {et~al.}(2010)\citenamefont
  {Roytershteyn}, \citenamefont {Daughton}, \citenamefont {Dorfman},
  \citenamefont {Ren}, \citenamefont {Ji}, \citenamefont {Yamada},
  \citenamefont {Karimabadi}, \citenamefont {Yin}, \citenamefont {Albright},\
  and\ \citenamefont {Bowers}}]{roytershteyn:2010}%
  \BibitemOpen
  \bibfield  {author} {\bibinfo {author} {\bibfnamefont {V.}~\bibnamefont
  {Roytershteyn}}, \bibinfo {author} {\bibfnamefont {W.}~\bibnamefont
  {Daughton}}, \bibinfo {author} {\bibfnamefont {S.}~\bibnamefont {Dorfman}},
  \bibinfo {author} {\bibfnamefont {Y.}~\bibnamefont {Ren}}, \bibinfo {author}
  {\bibfnamefont {H.}~\bibnamefont {Ji}}, \bibinfo {author} {\bibfnamefont
  {M.}~\bibnamefont {Yamada}}, \bibinfo {author} {\bibfnamefont
  {H.}~\bibnamefont {Karimabadi}}, \bibinfo {author} {\bibfnamefont
  {L.}~\bibnamefont {Yin}}, \bibinfo {author} {\bibfnamefont {B.}~\bibnamefont
  {Albright}}, \ and\ \bibinfo {author} {\bibfnamefont {K.}~\bibnamefont
  {Bowers}},\ }\href@noop {} {\bibfield  {journal} {\bibinfo  {journal}
  {Physics of Plasmas}\ }\textbf {\bibinfo {volume} {17}},\ \bibinfo {pages}
  {055706} (\bibinfo {year} {2010})}\BibitemShut {NoStop}%
\bibitem [{\citenamefont {Le}\ \emph {et~al.}(2015)\citenamefont {Le},
  \citenamefont {Egedal}, \citenamefont {Daughton}, \citenamefont
  {Roytershteyn}, \citenamefont {Karimabadi},\ and\ \citenamefont
  {Forest}}]{le:2015}%
  \BibitemOpen
  \bibfield  {author} {\bibinfo {author} {\bibfnamefont {A.}~\bibnamefont
  {Le}}, \bibinfo {author} {\bibfnamefont {J.}~\bibnamefont {Egedal}}, \bibinfo
  {author} {\bibfnamefont {W.}~\bibnamefont {Daughton}}, \bibinfo {author}
  {\bibfnamefont {V.}~\bibnamefont {Roytershteyn}}, \bibinfo {author}
  {\bibfnamefont {H.}~\bibnamefont {Karimabadi}}, \ and\ \bibinfo {author}
  {\bibfnamefont {C.}~\bibnamefont {Forest}},\ }\href@noop {} {\bibfield
  {journal} {\bibinfo  {journal} {Journal of Plasma Physics}\ }\textbf
  {\bibinfo {volume} {81}},\ \bibinfo {pages} {305810108} (\bibinfo {year}
  {2015})}\BibitemShut {NoStop}%
\bibitem [{\citenamefont {Higginson}, \citenamefont {Link},\ and\ \citenamefont
  {Schmidt}(2019)}]{higginson:2019}%
  \BibitemOpen
  \bibfield  {author} {\bibinfo {author} {\bibfnamefont {D.~P.}\ \bibnamefont
  {Higginson}}, \bibinfo {author} {\bibfnamefont {A.}~\bibnamefont {Link}}, \
  and\ \bibinfo {author} {\bibfnamefont {A.}~\bibnamefont {Schmidt}},\
  }\href@noop {} {\bibfield  {journal} {\bibinfo  {journal} {Journal of
  Computational Physics}\ }\textbf {\bibinfo {volume} {388}},\ \bibinfo {pages}
  {439} (\bibinfo {year} {2019})}\BibitemShut {NoStop}%
\bibitem [{\citenamefont {Lemons}\ \emph {et~al.}(2009)\citenamefont {Lemons},
  \citenamefont {Winske}, \citenamefont {Daughton},\ and\ \citenamefont
  {Albright}}]{lemons:2009}%
  \BibitemOpen
  \bibfield  {author} {\bibinfo {author} {\bibfnamefont {D.~S.}\ \bibnamefont
  {Lemons}}, \bibinfo {author} {\bibfnamefont {D.}~\bibnamefont {Winske}},
  \bibinfo {author} {\bibfnamefont {W.}~\bibnamefont {Daughton}}, \ and\
  \bibinfo {author} {\bibfnamefont {B.}~\bibnamefont {Albright}},\ }\href@noop
  {} {\bibfield  {journal} {\bibinfo  {journal} {Journal of Computational
  Physics}\ }\textbf {\bibinfo {volume} {228}},\ \bibinfo {pages} {1391}
  (\bibinfo {year} {2009})}\BibitemShut {NoStop}%
\bibitem [{\citenamefont {Molvig}\ \emph {et~al.}(2014)\citenamefont {Molvig},
  \citenamefont {Vold}, \citenamefont {Dodd},\ and\ \citenamefont
  {Wilks}}]{molvig:2014}%
  \BibitemOpen
  \bibfield  {author} {\bibinfo {author} {\bibfnamefont {K.}~\bibnamefont
  {Molvig}}, \bibinfo {author} {\bibfnamefont {E.~L.}\ \bibnamefont {Vold}},
  \bibinfo {author} {\bibfnamefont {E.~S.}\ \bibnamefont {Dodd}}, \ and\
  \bibinfo {author} {\bibfnamefont {S.~C.}\ \bibnamefont {Wilks}},\ }\href@noop
  {} {\bibfield  {journal} {\bibinfo  {journal} {Physical review letters}\
  }\textbf {\bibinfo {volume} {113}},\ \bibinfo {pages} {145001} (\bibinfo
  {year} {2014})}\BibitemShut {NoStop}%
\bibitem [{\citenamefont {Molvig}, \citenamefont {Simakov},\ and\ \citenamefont
  {Vold}(2014)}]{molvig:2014pop}%
  \BibitemOpen
  \bibfield  {author} {\bibinfo {author} {\bibfnamefont {K.}~\bibnamefont
  {Molvig}}, \bibinfo {author} {\bibfnamefont {A.~N.}\ \bibnamefont {Simakov}},
  \ and\ \bibinfo {author} {\bibfnamefont {E.~L.}\ \bibnamefont {Vold}},\
  }\href@noop {} {\bibfield  {journal} {\bibinfo  {journal} {Physics of
  Plasmas}\ }\textbf {\bibinfo {volume} {21}},\ \bibinfo {pages} {092709}
  (\bibinfo {year} {2014})}\BibitemShut {NoStop}%
\bibitem [{\citenamefont {Haines}\ \emph {et~al.}(2014)\citenamefont {Haines},
  \citenamefont {Vold}, \citenamefont {Molvig}, \citenamefont {Aldrich},\ and\
  \citenamefont {Rauenzahn}}]{haines:2014}%
  \BibitemOpen
  \bibfield  {author} {\bibinfo {author} {\bibfnamefont {B.~M.}\ \bibnamefont
  {Haines}}, \bibinfo {author} {\bibfnamefont {E.~L.}\ \bibnamefont {Vold}},
  \bibinfo {author} {\bibfnamefont {K.}~\bibnamefont {Molvig}}, \bibinfo
  {author} {\bibfnamefont {C.}~\bibnamefont {Aldrich}}, \ and\ \bibinfo
  {author} {\bibfnamefont {R.}~\bibnamefont {Rauenzahn}},\ }\href@noop {}
  {\bibfield  {journal} {\bibinfo  {journal} {Physics of Plasmas}\ }\textbf
  {\bibinfo {volume} {21}},\ \bibinfo {pages} {092306} (\bibinfo {year}
  {2014})}\BibitemShut {NoStop}%
\bibitem [{\citenamefont {Simakov}\ and\ \citenamefont
  {Molvig}(2016{\natexlab{a}})}]{simakov:2016}%
  \BibitemOpen
  \bibfield  {author} {\bibinfo {author} {\bibfnamefont {A.~N.}\ \bibnamefont
  {Simakov}}\ and\ \bibinfo {author} {\bibfnamefont {K.}~\bibnamefont
  {Molvig}},\ }\href@noop {} {\bibfield  {journal} {\bibinfo  {journal}
  {Physics of Plasmas}\ }\textbf {\bibinfo {volume} {23}},\ \bibinfo {pages}
  {032115} (\bibinfo {year} {2016}{\natexlab{a}})}\BibitemShut {NoStop}%
\bibitem [{\citenamefont {Simakov}\ and\ \citenamefont
  {Molvig}(2016{\natexlab{b}})}]{simakov:2016b}%
  \BibitemOpen
  \bibfield  {author} {\bibinfo {author} {\bibfnamefont {A.~N.}\ \bibnamefont
  {Simakov}}\ and\ \bibinfo {author} {\bibfnamefont {K.}~\bibnamefont
  {Molvig}},\ }\href@noop {} {\bibfield  {journal} {\bibinfo  {journal}
  {Physics of Plasmas}\ }\textbf {\bibinfo {volume} {23}},\ \bibinfo {pages}
  {032116} (\bibinfo {year} {2016}{\natexlab{b}})}\BibitemShut {NoStop}%
\bibitem [{\citenamefont {Vold}\ \emph {et~al.}(2017)\citenamefont {Vold},
  \citenamefont {Rauenzahn}, \citenamefont {Aldrich}, \citenamefont {Molvig},
  \citenamefont {Simakov},\ and\ \citenamefont {Haines}}]{vold:2017}%
  \BibitemOpen
  \bibfield  {author} {\bibinfo {author} {\bibfnamefont {E.~L.}\ \bibnamefont
  {Vold}}, \bibinfo {author} {\bibfnamefont {R.~M.}\ \bibnamefont {Rauenzahn}},
  \bibinfo {author} {\bibfnamefont {C.}~\bibnamefont {Aldrich}}, \bibinfo
  {author} {\bibfnamefont {K.}~\bibnamefont {Molvig}}, \bibinfo {author}
  {\bibfnamefont {A.~N.}\ \bibnamefont {Simakov}}, \ and\ \bibinfo {author}
  {\bibfnamefont {B.~M.}\ \bibnamefont {Haines}},\ }\href@noop {} {\bibfield
  {journal} {\bibinfo  {journal} {Physics of Plasmas}\ }\textbf {\bibinfo
  {volume} {24}},\ \bibinfo {pages} {042702} (\bibinfo {year}
  {2017})}\BibitemShut {NoStop}%
\bibitem [{\citenamefont {Vold}\ \emph
  {et~al.}(2018{\natexlab{a}})\citenamefont {Vold}, \citenamefont {Yin},
  \citenamefont {Taitano}, \citenamefont {Molvig},\ and\ \citenamefont
  {Albright}}]{vold:2018}%
  \BibitemOpen
  \bibfield  {author} {\bibinfo {author} {\bibfnamefont {E.~L.}\ \bibnamefont
  {Vold}}, \bibinfo {author} {\bibfnamefont {L.}~\bibnamefont {Yin}}, \bibinfo
  {author} {\bibfnamefont {W.}~\bibnamefont {Taitano}}, \bibinfo {author}
  {\bibfnamefont {K.}~\bibnamefont {Molvig}}, \ and\ \bibinfo {author}
  {\bibfnamefont {B.~J.}\ \bibnamefont {Albright}},\ }\href@noop {} {\bibfield
  {journal} {\bibinfo  {journal} {Physics of Plasmas}\ }\textbf {\bibinfo
  {volume} {25}},\ \bibinfo {pages} {062102} (\bibinfo {year}
  {2018}{\natexlab{a}})}\BibitemShut {NoStop}%
\bibitem [{\citenamefont {Vold}\ \emph
  {et~al.}(2018{\natexlab{b}})\citenamefont {Vold}, \citenamefont {Kagan},
  \citenamefont {Simakov}, \citenamefont {Molvig},\ and\ \citenamefont
  {Yin}}]{vold:2018b}%
  \BibitemOpen
  \bibfield  {author} {\bibinfo {author} {\bibfnamefont {E.}~\bibnamefont
  {Vold}}, \bibinfo {author} {\bibfnamefont {G.}~\bibnamefont {Kagan}},
  \bibinfo {author} {\bibfnamefont {A.~N.}\ \bibnamefont {Simakov}}, \bibinfo
  {author} {\bibfnamefont {K.}~\bibnamefont {Molvig}}, \ and\ \bibinfo {author}
  {\bibfnamefont {L.}~\bibnamefont {Yin}},\ }\href@noop {} {\bibfield
  {journal} {\bibinfo  {journal} {Plasma Physics and Controlled Fusion}\
  }\textbf {\bibinfo {volume} {60}},\ \bibinfo {pages} {054010} (\bibinfo
  {year} {2018}{\natexlab{b}})}\BibitemShut {NoStop}%
\bibitem [{\citenamefont {Yin}\ \emph {et~al.}(2016)\citenamefont {Yin},
  \citenamefont {Albright}, \citenamefont {Taitano}, \citenamefont {Vold},
  \citenamefont {Chacon},\ and\ \citenamefont {Simakov}}]{yin:2016}%
  \BibitemOpen
  \bibfield  {author} {\bibinfo {author} {\bibfnamefont {L.}~\bibnamefont
  {Yin}}, \bibinfo {author} {\bibfnamefont {B.}~\bibnamefont {Albright}},
  \bibinfo {author} {\bibfnamefont {W.}~\bibnamefont {Taitano}}, \bibinfo
  {author} {\bibfnamefont {E.}~\bibnamefont {Vold}}, \bibinfo {author}
  {\bibfnamefont {L.}~\bibnamefont {Chacon}}, \ and\ \bibinfo {author}
  {\bibfnamefont {A.}~\bibnamefont {Simakov}},\ }\href@noop {} {\bibfield
  {journal} {\bibinfo  {journal} {Physics of Plasmas}\ }\textbf {\bibinfo
  {volume} {23}},\ \bibinfo {pages} {112302} (\bibinfo {year}
  {2016})}\BibitemShut {NoStop}%
\bibitem [{\citenamefont {Yin}\ \emph {et~al.}(2019)\citenamefont {Yin},
  \citenamefont {Albright}, \citenamefont {Vold}, \citenamefont {Nystrom},
  \citenamefont {Bird},\ and\ \citenamefont {Bowers}}]{yin:2019}%
  \BibitemOpen
  \bibfield  {author} {\bibinfo {author} {\bibfnamefont {L.}~\bibnamefont
  {Yin}}, \bibinfo {author} {\bibfnamefont {B.~J.}\ \bibnamefont {Albright}},
  \bibinfo {author} {\bibfnamefont {E.~L.}\ \bibnamefont {Vold}}, \bibinfo
  {author} {\bibfnamefont {W.~D.}\ \bibnamefont {Nystrom}}, \bibinfo {author}
  {\bibfnamefont {R.~F.}\ \bibnamefont {Bird}}, \ and\ \bibinfo {author}
  {\bibfnamefont {K.~J.}\ \bibnamefont {Bowers}},\ }\href@noop {} {\bibfield
  {journal} {\bibinfo  {journal} {Physics of Plasmas}\ }\textbf {\bibinfo
  {volume} {26}},\ \bibinfo {pages} {062302} (\bibinfo {year}
  {2019})}\BibitemShut {NoStop}%
\bibitem [{\citenamefont {Vold}, \citenamefont {Yin},\ and\ \citenamefont
  {Albright}(2021)}]{vold:2021}%
  \BibitemOpen
  \bibfield  {author} {\bibinfo {author} {\bibfnamefont {E.}~\bibnamefont
  {Vold}}, \bibinfo {author} {\bibfnamefont {L.}~\bibnamefont {Yin}}, \ and\
  \bibinfo {author} {\bibfnamefont {B.}~\bibnamefont {Albright}},\ }\href@noop
  {} {\bibfield  {journal} {\bibinfo  {journal} {Physics of Plasmas}\ }\textbf
  {\bibinfo {volume} {28}},\ \bibinfo {pages} {092709} (\bibinfo {year}
  {2021})}\BibitemShut {NoStop}%
\bibitem [{\citenamefont {Bird}\ \emph {et~al.}(2021)\citenamefont {Bird},
  \citenamefont {Tan}, \citenamefont {Luedtke}, \citenamefont {Harrell},
  \citenamefont {Taufer},\ and\ \citenamefont {Albright}}]{bird:2021}%
  \BibitemOpen
  \bibfield  {author} {\bibinfo {author} {\bibfnamefont {R.}~\bibnamefont
  {Bird}}, \bibinfo {author} {\bibfnamefont {N.}~\bibnamefont {Tan}}, \bibinfo
  {author} {\bibfnamefont {S.~V.}\ \bibnamefont {Luedtke}}, \bibinfo {author}
  {\bibfnamefont {S.~L.}\ \bibnamefont {Harrell}}, \bibinfo {author}
  {\bibfnamefont {M.}~\bibnamefont {Taufer}}, \ and\ \bibinfo {author}
  {\bibfnamefont {B.}~\bibnamefont {Albright}},\ }\href@noop {} {\bibfield
  {journal} {\bibinfo  {journal} {IEEE Transactions on Parallel and Distributed
  Systems}\ }\textbf {\bibinfo {volume} {33}},\ \bibinfo {pages} {952}
  (\bibinfo {year} {2021})}\BibitemShut {NoStop}%
\bibitem [{\citenamefont {Gittings}\ \emph {et~al.}(2008)\citenamefont
  {Gittings}, \citenamefont {Weaver}, \citenamefont {Clover}, \citenamefont
  {Betlach}, \citenamefont {Byrne}, \citenamefont {Coker}, \citenamefont
  {Dendy}, \citenamefont {Hueckstaedt}, \citenamefont {New}, \citenamefont
  {Oakes}, \citenamefont {Ranta},\ and\ \citenamefont
  {Stefan}}]{gittings:2008}%
  \BibitemOpen
  \bibfield  {author} {\bibinfo {author} {\bibfnamefont {M.}~\bibnamefont
  {Gittings}}, \bibinfo {author} {\bibfnamefont {R.}~\bibnamefont {Weaver}},
  \bibinfo {author} {\bibfnamefont {M.}~\bibnamefont {Clover}}, \bibinfo
  {author} {\bibfnamefont {T.}~\bibnamefont {Betlach}}, \bibinfo {author}
  {\bibfnamefont {N.}~\bibnamefont {Byrne}}, \bibinfo {author} {\bibfnamefont
  {R.}~\bibnamefont {Coker}}, \bibinfo {author} {\bibfnamefont
  {E.}~\bibnamefont {Dendy}}, \bibinfo {author} {\bibfnamefont
  {R.}~\bibnamefont {Hueckstaedt}}, \bibinfo {author} {\bibfnamefont
  {K.}~\bibnamefont {New}}, \bibinfo {author} {\bibfnamefont {W.~R.}\
  \bibnamefont {Oakes}}, \bibinfo {author} {\bibfnamefont {D.}~\bibnamefont
  {Ranta}}, \ and\ \bibinfo {author} {\bibfnamefont {R.}~\bibnamefont
  {Stefan}},\ }\href@noop {} {\bibfield  {journal} {\bibinfo  {journal}
  {Comput. Sci. Discovery}\ }\textbf {\bibinfo {volume} {1}} (\bibinfo {year}
  {2008})}\BibitemShut {NoStop}%
\bibitem [{\citenamefont {Haines}\ \emph {et~al.}(2017)\citenamefont {Haines},
  \citenamefont {Aldrich}, \citenamefont {Campbell}, \citenamefont
  {Rauenzahn},\ and\ \citenamefont {Wingate}}]{haines:2017}%
  \BibitemOpen
  \bibfield  {author} {\bibinfo {author} {\bibfnamefont {B.~M.}\ \bibnamefont
  {Haines}}, \bibinfo {author} {\bibfnamefont {C.~H.}\ \bibnamefont {Aldrich}},
  \bibinfo {author} {\bibfnamefont {J.~M.}\ \bibnamefont {Campbell}}, \bibinfo
  {author} {\bibfnamefont {R.~M.}\ \bibnamefont {Rauenzahn}}, \ and\ \bibinfo
  {author} {\bibfnamefont {C.~A.}\ \bibnamefont {Wingate}},\ }\href@noop {}
  {\bibfield  {journal} {\bibinfo  {journal} {Physics of Plasmas}\ }\textbf
  {\bibinfo {volume} {24}} (\bibinfo {year} {2017})}\BibitemShut {NoStop}%
\bibitem [{\citenamefont {Besnard}\ \emph {et~al.}(1992)\citenamefont
  {Besnard}, \citenamefont {Harlow}, \citenamefont {Rauenzahn},\ and\
  \citenamefont {Zemach}}]{besnard:1992}%
  \BibitemOpen
  \bibfield  {author} {\bibinfo {author} {\bibfnamefont {D.}~\bibnamefont
  {Besnard}}, \bibinfo {author} {\bibfnamefont {F.~H.}\ \bibnamefont {Harlow}},
  \bibinfo {author} {\bibfnamefont {R.~M.}\ \bibnamefont {Rauenzahn}}, \ and\
  \bibinfo {author} {\bibfnamefont {C.}~\bibnamefont {Zemach}},\ }\href@noop {}
  {\enquote {\bibinfo {title} {Turbulence transport equations for
  variable-density turbulence and their relationship to two-field models},}\
  }\bibinfo {type} {Tech. Rep.}\ (\bibinfo  {institution} {Los Alamos National
  Lab., NM (United States)},\ \bibinfo {year} {1992})\BibitemShut {NoStop}%
\bibitem [{\citenamefont {Banerjee}, \citenamefont {Gore},\ and\ \citenamefont
  {Andrews}(2010)}]{banerjee:2010}%
  \BibitemOpen
  \bibfield  {author} {\bibinfo {author} {\bibfnamefont {A.}~\bibnamefont
  {Banerjee}}, \bibinfo {author} {\bibfnamefont {R.~A.}\ \bibnamefont {Gore}},
  \ and\ \bibinfo {author} {\bibfnamefont {M.~J.}\ \bibnamefont {Andrews}},\
  }\href@noop {} {\bibfield  {journal} {\bibinfo  {journal} {Physical Review
  E}\ }\textbf {\bibinfo {volume} {82}},\ \bibinfo {pages} {046309} (\bibinfo
  {year} {2010})}\BibitemShut {NoStop}%
\bibitem [{\citenamefont {Ristorcelli}(2017)}]{ristorcelli:2017}%
  \BibitemOpen
  \bibfield  {author} {\bibinfo {author} {\bibfnamefont {J.~R.}\ \bibnamefont
  {Ristorcelli}},\ }\href@noop {} {\bibfield  {journal} {\bibinfo  {journal}
  {Physics of Fluids}\ }\textbf {\bibinfo {volume} {29}},\ \bibinfo {pages}
  {020705} (\bibinfo {year} {2017})}\BibitemShut {NoStop}%
\bibitem [{\citenamefont {Albright}\ \emph {et~al.}(2022)\citenamefont
  {Albright}, \citenamefont {Murphy}, \citenamefont {Haines}, \citenamefont
  {Douglas}, \citenamefont {Cooley}, \citenamefont {Day}, \citenamefont
  {Denissen}, \citenamefont {Di~Stefano}, \citenamefont {Donovan},
  \citenamefont {Edwards} \emph {et~al.}}]{albright:2022}%
  \BibitemOpen
  \bibfield  {author} {\bibinfo {author} {\bibfnamefont {B.~J.}\ \bibnamefont
  {Albright}}, \bibinfo {author} {\bibfnamefont {T.~J.}\ \bibnamefont
  {Murphy}}, \bibinfo {author} {\bibfnamefont {B.}~\bibnamefont {Haines}},
  \bibinfo {author} {\bibfnamefont {M.}~\bibnamefont {Douglas}}, \bibinfo
  {author} {\bibfnamefont {J.~H.}\ \bibnamefont {Cooley}}, \bibinfo {author}
  {\bibfnamefont {T.~H.}\ \bibnamefont {Day}}, \bibinfo {author} {\bibfnamefont
  {N.~A.}\ \bibnamefont {Denissen}}, \bibinfo {author} {\bibfnamefont
  {C.}~\bibnamefont {Di~Stefano}}, \bibinfo {author} {\bibfnamefont
  {P.}~\bibnamefont {Donovan}}, \bibinfo {author} {\bibfnamefont
  {S.}~\bibnamefont {Edwards}},  \emph {et~al.},\ }\href@noop {} {\bibfield
  {journal} {\bibinfo  {journal} {Physics of Plasmas}\ }\textbf {\bibinfo
  {volume} {29}},\ \bibinfo {pages} {022702} (\bibinfo {year}
  {2022})}\BibitemShut {NoStop}%
\bibitem [{\citenamefont {Robey}\ \emph {et~al.}(2003)\citenamefont {Robey},
  \citenamefont {Zhou}, \citenamefont {Buckingham}, \citenamefont {Keiter},
  \citenamefont {Remington},\ and\ \citenamefont {Drake}}]{robey:2003}%
  \BibitemOpen
  \bibfield  {author} {\bibinfo {author} {\bibfnamefont {H.}~\bibnamefont
  {Robey}}, \bibinfo {author} {\bibfnamefont {Y.}~\bibnamefont {Zhou}},
  \bibinfo {author} {\bibfnamefont {A.}~\bibnamefont {Buckingham}}, \bibinfo
  {author} {\bibfnamefont {P.}~\bibnamefont {Keiter}}, \bibinfo {author}
  {\bibfnamefont {B.~A.}\ \bibnamefont {Remington}}, \ and\ \bibinfo {author}
  {\bibfnamefont {R.~P.}\ \bibnamefont {Drake}},\ }\href@noop {} {\bibfield
  {journal} {\bibinfo  {journal} {Physics of Plasmas}\ }\textbf {\bibinfo
  {volume} {10}},\ \bibinfo {pages} {614} (\bibinfo {year} {2003})}\BibitemShut
  {NoStop}%
\bibitem [{\citenamefont {Weber}\ \emph {et~al.}(2014)\citenamefont {Weber},
  \citenamefont {Clark}, \citenamefont {Cook}, \citenamefont {Busby},\ and\
  \citenamefont {Robey}}]{weber:2014}%
  \BibitemOpen
  \bibfield  {author} {\bibinfo {author} {\bibfnamefont {C.}~\bibnamefont
  {Weber}}, \bibinfo {author} {\bibfnamefont {D.}~\bibnamefont {Clark}},
  \bibinfo {author} {\bibfnamefont {A.}~\bibnamefont {Cook}}, \bibinfo {author}
  {\bibfnamefont {L.}~\bibnamefont {Busby}}, \ and\ \bibinfo {author}
  {\bibfnamefont {H.}~\bibnamefont {Robey}},\ }\href@noop {} {\bibfield
  {journal} {\bibinfo  {journal} {Physical Review E}\ }\textbf {\bibinfo
  {volume} {89}},\ \bibinfo {pages} {053106} (\bibinfo {year}
  {2014})}\BibitemShut {NoStop}%
\bibitem [{\citenamefont {Weber}\ \emph {et~al.}(2015)\citenamefont {Weber},
  \citenamefont {Clark}, \citenamefont {Cook}, \citenamefont {Eder},
  \citenamefont {Haan}, \citenamefont {Hammel}, \citenamefont {Hinkel},
  \citenamefont {Jones}, \citenamefont {Marinak}, \citenamefont {Milovich}
  \emph {et~al.}}]{weber:2015}%
  \BibitemOpen
  \bibfield  {author} {\bibinfo {author} {\bibfnamefont {C.}~\bibnamefont
  {Weber}}, \bibinfo {author} {\bibfnamefont {D.}~\bibnamefont {Clark}},
  \bibinfo {author} {\bibfnamefont {A.}~\bibnamefont {Cook}}, \bibinfo {author}
  {\bibfnamefont {D.}~\bibnamefont {Eder}}, \bibinfo {author} {\bibfnamefont
  {S.}~\bibnamefont {Haan}}, \bibinfo {author} {\bibfnamefont {B.}~\bibnamefont
  {Hammel}}, \bibinfo {author} {\bibfnamefont {D.}~\bibnamefont {Hinkel}},
  \bibinfo {author} {\bibfnamefont {O.}~\bibnamefont {Jones}}, \bibinfo
  {author} {\bibfnamefont {M.}~\bibnamefont {Marinak}}, \bibinfo {author}
  {\bibfnamefont {J.}~\bibnamefont {Milovich}},  \emph {et~al.},\ }\href@noop
  {} {\bibfield  {journal} {\bibinfo  {journal} {Physics of Plasmas}\ }\textbf
  {\bibinfo {volume} {22}},\ \bibinfo {pages} {032702} (\bibinfo {year}
  {2015})}\BibitemShut {NoStop}%
\bibitem [{\citenamefont {Abu-Shawareb}\ \emph {et~al.}(2022)\citenamefont
  {Abu-Shawareb}, \citenamefont {Acree}, \citenamefont {Adams}, \citenamefont
  {Adams}, \citenamefont {Addis}, \citenamefont {Aden}, \citenamefont {Adrian},
  \citenamefont {Afeyan}, \citenamefont {Aggleton}, \citenamefont {Aghaian}
  \emph {et~al.}}]{abu:2022}%
  \BibitemOpen
  \bibfield  {author} {\bibinfo {author} {\bibfnamefont {H.}~\bibnamefont
  {Abu-Shawareb}}, \bibinfo {author} {\bibfnamefont {R.}~\bibnamefont {Acree}},
  \bibinfo {author} {\bibfnamefont {P.}~\bibnamefont {Adams}}, \bibinfo
  {author} {\bibfnamefont {J.}~\bibnamefont {Adams}}, \bibinfo {author}
  {\bibfnamefont {B.}~\bibnamefont {Addis}}, \bibinfo {author} {\bibfnamefont
  {R.}~\bibnamefont {Aden}}, \bibinfo {author} {\bibfnamefont {P.}~\bibnamefont
  {Adrian}}, \bibinfo {author} {\bibfnamefont {B.}~\bibnamefont {Afeyan}},
  \bibinfo {author} {\bibfnamefont {M.}~\bibnamefont {Aggleton}}, \bibinfo
  {author} {\bibfnamefont {L.}~\bibnamefont {Aghaian}},  \emph {et~al.},\
  }\href@noop {} {\bibfield  {journal} {\bibinfo  {journal} {Physical Review
  Letters}\ }\textbf {\bibinfo {volume} {129}},\ \bibinfo {pages} {075001}
  (\bibinfo {year} {2022})}\BibitemShut {NoStop}%
\bibitem [{\citenamefont {Daughton}\ \emph {et~al.}(2023)\citenamefont
  {Daughton}, \citenamefont {Albright}, \citenamefont {Finnegan}, \citenamefont
  {Haines}, \citenamefont {Kline}, \citenamefont {Sauppe},\ and\ \citenamefont
  {Smidt}}]{daughton:2023}%
  \BibitemOpen
  \bibfield  {author} {\bibinfo {author} {\bibfnamefont {W.}~\bibnamefont
  {Daughton}}, \bibinfo {author} {\bibfnamefont {B.}~\bibnamefont {Albright}},
  \bibinfo {author} {\bibfnamefont {S.}~\bibnamefont {Finnegan}}, \bibinfo
  {author} {\bibfnamefont {B.~M.}\ \bibnamefont {Haines}}, \bibinfo {author}
  {\bibfnamefont {J.}~\bibnamefont {Kline}}, \bibinfo {author} {\bibfnamefont
  {J.}~\bibnamefont {Sauppe}}, \ and\ \bibinfo {author} {\bibfnamefont
  {J.}~\bibnamefont {Smidt}},\ }\href@noop {} {\bibfield  {journal} {\bibinfo
  {journal} {Physics of Plasmas}\ }\textbf {\bibinfo {volume} {30}},\ \bibinfo
  {pages} {012704} (\bibinfo {year} {2023})}\BibitemShut {NoStop}%
\bibitem [{\citenamefont {Huba}(1998)}]{huba:1998}%
  \BibitemOpen
  \bibfield  {author} {\bibinfo {author} {\bibfnamefont {J.~D.}\ \bibnamefont
  {Huba}},\ }\href@noop {} {\emph {\bibinfo {title} {NRL plasma formulary}}},\
  Vol.\ \bibinfo {volume} {6790}\ (\bibinfo  {publisher} {Naval Research
  Laboratory},\ \bibinfo {year} {1998})\BibitemShut {NoStop}%
\bibitem [{\citenamefont {Higginson}\ and\ \citenamefont
  {Link}(2022)}]{higginson:2022}%
  \BibitemOpen
  \bibfield  {author} {\bibinfo {author} {\bibfnamefont {D.~P.}\ \bibnamefont
  {Higginson}}\ and\ \bibinfo {author} {\bibfnamefont {A.~J.}\ \bibnamefont
  {Link}},\ }\href@noop {} {\bibfield  {journal} {\bibinfo  {journal} {Journal
  of Computational Physics}\ }\textbf {\bibinfo {volume} {457}},\ \bibinfo
  {pages} {110935} (\bibinfo {year} {2022})}\BibitemShut {NoStop}%
\bibitem [{\citenamefont {Simakov}(2022)}]{simakov:2022}%
  \BibitemOpen
  \bibfield  {author} {\bibinfo {author} {\bibfnamefont {A.~N.}\ \bibnamefont
  {Simakov}},\ }\href@noop {} {\bibfield  {journal} {\bibinfo  {journal}
  {Physics of Plasmas}\ }\textbf {\bibinfo {volume} {29}},\ \bibinfo {pages}
  {022304} (\bibinfo {year} {2022})}\BibitemShut {NoStop}%
\bibitem [{\citenamefont {Braginskii}(1958)}]{braginskii:1958}%
  \BibitemOpen
  \bibfield  {author} {\bibinfo {author} {\bibfnamefont {S.}~\bibnamefont
  {Braginskii}},\ }\href@noop {} {\bibfield  {journal} {\bibinfo  {journal}
  {Sov. Phys. JETP}\ }\textbf {\bibinfo {volume} {6}},\ \bibinfo {pages} {358}
  (\bibinfo {year} {1958})}\BibitemShut {NoStop}%
\bibitem [{\citenamefont {Edwards}, \citenamefont {Trott},\ and\ \citenamefont
  {Sunderland}(2014)}]{edwards:2014}%
  \BibitemOpen
  \bibfield  {author} {\bibinfo {author} {\bibfnamefont {H.~C.}\ \bibnamefont
  {Edwards}}, \bibinfo {author} {\bibfnamefont {C.~R.}\ \bibnamefont {Trott}},
  \ and\ \bibinfo {author} {\bibfnamefont {D.}~\bibnamefont {Sunderland}},\
  }\href@noop {} {\bibfield  {journal} {\bibinfo  {journal} {Journal of
  parallel and distributed computing}\ }\textbf {\bibinfo {volume} {74}},\
  \bibinfo {pages} {3202} (\bibinfo {year} {2014})}\BibitemShut {NoStop}%
\bibitem [{\citenamefont {Rosenbluth}(1960)}]{rosenbluth:1960}%
  \BibitemOpen
  \bibfield  {author} {\bibinfo {author} {\bibfnamefont {M.~N.}\ \bibnamefont
  {Rosenbluth}},\ }\href@noop {} {\bibfield  {journal} {\bibinfo  {journal}
  {The Physics of Fluids}\ }\textbf {\bibinfo {volume} {3}},\ \bibinfo {pages}
  {932} (\bibinfo {year} {1960})}\BibitemShut {NoStop}%
\bibitem [{\citenamefont {Uhm}\ and\ \citenamefont {Lampe}(1980)}]{uhm:1980}%
  \BibitemOpen
  \bibfield  {author} {\bibinfo {author} {\bibfnamefont {H.~S.}\ \bibnamefont
  {Uhm}}\ and\ \bibinfo {author} {\bibfnamefont {M.}~\bibnamefont {Lampe}},\
  }\href@noop {} {\bibfield  {journal} {\bibinfo  {journal} {The Physics of
  Fluids}\ }\textbf {\bibinfo {volume} {23}},\ \bibinfo {pages} {1574}
  (\bibinfo {year} {1980})}\BibitemShut {NoStop}%
\end{thebibliography}

%

\end{document}